\pdfoutput=1
\documentclass[11pt,twoside]{article}
\usepackage[backend=biber]{biblatex}
\bibliography{On_The_Equivalence_Of_The_Mean_Variance_Criterion_And_Stochastic_Dominance_Criteria.bib}
\usepackage[utf8]{inputenc}
\usepackage{mathtools}
\usepackage{etoolbox}
\usepackage{enumitem}
\usepackage{adjustbox}
\usepackage{graphics}
\usepackage{fancyhdr}
\usepackage[toc,page]{appendix}
\makeatletter
\usepackage{ctable}
\usepackage[flushleft]{threeparttable}
\usepackage{titletoc}
\usepackage{fix-cm}
\usepackage{datetime}
\usepackage{ragged2e}
\usepackage{lipsum}
\usepackage{subcaption}
\usepackage{float}
\usepackage{footnote}
\usepackage{amsmath, amssymb, amsthm} 
\usepackage{graphicx,color}           
\usepackage[left=1in, right=1in, top=1in, bottom=1in, includefoot]{geometry}
\usepackage{booktabs,caption}
\usepackage{dsfont}
\usepackage{multirow,array}
\usepackage{xcolor}
\usepackage{titlesec}

\newtheorem{theorem}{Theorem}[section]
\numberwithin{theorem}{section}

\newtheorem{condition}[theorem]{Condition}

\newtheorem{corollary}[theorem]{Corollary}

\newtheorem{definition}[theorem]{Definition}

\newtheorem{proposition}[theorem]{Proposition}
\newtheorem{remark}[theorem]{Remark}

\DeclareFieldFormat[article, inbook]{title}{#1}

\title{On The Equivalence Of The Mean Variance Criterion And Stochastic Dominance Criteria}

\date{\today}

\author{ George Samartzis\\
	Department of Banking and Financial Management\\
	University of Piraeus\\
	Piraeus, 18534 \\
	\texttt{gsamartzis@unipi.gr} \\
	\\
	
	Nikitas Pittis\\
	Department of Banking and Financial Management\\
	University of Piraeus\\
	Piraeus, 18534 \\
	\texttt{npittis@unipi.gr} \\
}

\providecommand{\keywords}[1]{\textbf{\textit{Index terms }} #1}

\begin{document}
	\maketitle
	
	\begin{abstract}
		We study the necessary and sufficient conditions under which the Mean-Variance Criterion (MVC) is equivalent to the Maximum Expected Utility Criterion (MEUC), for two lotteries. Based on Chamberlain (1983), we conclude that the MVC is equivalent to the Second-order Stochastic Dominance Rule (SSDR) under any symmetric Elliptical distribution. We then discuss the work of Schuhmacher et al. (2021). Although their theoretical findings deduce that the Mean-Variance Analysis remains valid under Skew-Elliptical distributions, we argue that this does not entail that the MVC coincides with the SSDR. In fact, generating multiple MV-pairs that follow a Skew-Normal distribution it becomes evident that the MVC fails to coincide with the SSDR for some types of risk-averse investors. In the second part of this work, we examine the premise of Levy and Markowitz (1979) that "the MVC deduces the maximization of the expected utility of an investor, under any approximately quadratic utility function, without making any further assumption on the distribution of the lotteries". Using Monte Carlo Simulations, we find out that the set of approximately quadratic utility functions is too narrow. Specifically, our simulations indicate that $\log{(a+Z)}$ and $(1+Z)^a$ are almost quadratic, while $-e^{-a(1+Z)}$ and $-(1+Z)^{-a}$ fail to approximate a quadratic utility function under either an Extreme Value or a Stable Pareto distribution.
	\end{abstract}

	\keywords{MVC, \and MEUC, \and SD, \and SDR}

	\section{Introduction}
	Back in the $18^{th}$ century Daniel Bernoulli proposed a solution for the St. Petersburg paradox. The solution was simply based on the assumption that an investor aims at maximizing his expected utility rather than his wealth. The utility function had to be logarithmic, meaning that it only covered the case where investors are risk averse. His most important assumption was that utility is both normative and descriptive. This simply means that "\textit{An investor not only is obliged to choose between different goods to maximize his expected utility but also does so in reality}". This premise did not however explain why people choose to gamble. So, the main issue with Bernoulli's approach was it lacked generality.
	
	Despite the fact that this finding was so important from a theoretical point of view it was not until 1944 that Von Neumann and Morgenstern (hereafter; VN-M) reinstated the theory around expected utility \cite{Neumann_Morgenstern}. In fact, they laid down a set of sufficient and necessary axioms that the preferences of a decision maker need to satisfy in order to conclude that he makes his decisions based on the maximization of his expected utility. The utility theory is now well-established and covers all different types of risk attitude. Until this day, the Maximum Expected Utility criterion (hereafter; MEUC) is considered the new canon of economic theory.
	
	However, in reality, we cannot apply direct utility maximization of an investor. The reason is that it is almost impossible to be aware of his exact utility function. What we could hope for would be to know his risk appetite (e.g. risk-averse). This is where the Stochastic Dominance (hereafter; SD) rules come to place. These rules constitute well-defined Theorems determining the necessary and sufficient conditions under which an investor with specific risk preferences maximizes his expected utility. These conditions have to do with specific properties of the distribution of lotteries.
	
	What we are going to focus more on is a different rule called the Mean Variance Criterion (hereafter; MVC) developed by Markowitz \cite{Markowitz1952a}, \cite{Markowitz1959} (1952, 1959). This rule is a decision making criterion based only on the first two moments of the distributions of two lotteries. We shall discuss its value as well as the root causes it has been criticized, like in Gandhi (1981) \cite{Gandhi1981}. The main source of interest is: how this criterion differs from SD rules and when should an investor use it. We will see that there are some widespread misunderstandings concerning the necessary and sufficient conditions under which MVC coincides with MEUC. For example, there is a general agreement in the literature that the MVC is meaningful under two scenarios: (i) either lotteries are normally distributed, or (i) the investor has quadratic utility. As Baron (1977) \cite{Baron1977} states, these assumptions are sufficient to justify the use of mean variance analysis in a manner consistent with the von Neumann-Morgenstern axioms. Delving into the theory and the vast existing literature, we will see that there are misconceptions regarding both of these assumptions which need to be discussed. In particular, as we shall see there are also more interesting assumptions that make the MVC coincide with the MEUC that we will thoroughly analyze. Then, we are going to focus on the main subject of this work which has to do with the following idea from Levy and Markowitz (1979) \cite{Levy1979} "Assuming we have an approximately quadratic utility function, MV choices will almost maximize our expected utility function". We will revisit this premise and we will try to give our view based on Monte Carlo simulations and a thorough analysis of the idea. Although this subject is relatively old, the  academic interest is still vivid. To name but a few of the most recent works, Markowitz (2010,2014) \cite{Markowitz2010}, \cite{Markowitz2014}, Bodnar et al. (2020) \cite{Bodnar2020}, Malavasi et al. (2020) \cite{Malavasi2020} and Schuhmacher et al. (2021) \cite{Schuhmacher2021}.
	
	In the following sections, we are going to approach this subject carefully so as to not leave any unanswered questions. We start by setting the theoretical framework as constructed by VN-M. Based on it, we will define the MVC as well as the SD rules. From there on, we will delve into the necessary and sufficient conditions that connect the MVC with the SD rules and hence with the MEUC. We will study the assumptions of either normality or quadratic utility independently and discuss whether or not they are very restrictive. Next, we will discuss the more interesting cases of Elliptical and Skew-Elliptical families of distribution that seem to elevate the value of the MVC. Finally, we will examine the premise of Levy and Markowitz using Monte Carlo simulations together with a careful connection to the findings of the existing literature.
	
	Our main findings split into two parts. (i) We prove that, contrary to Elliptical distributions, under Skew-Elliptical distributions the MVC is not optimal for any risk-averse investor. In fact, our Monte Carlo simulations derive cases under which the MVC fails to deduce the right decision. (ii) The premise of Markowitz with respect to the approximately quadratic utility functions seems to be valid for Skew-Elliptical distributions. As we deviate more from normality (e.g. Extreme Value and Stable Pareto distributions) the set of approximately quadratic utility functions shrinks.
	
	\section{Theoretical Framework}
	An investor is given a number of different lotteries to choose from, each one with its own distribution. In order for the investor to make a rational choice his preferences should satisfy some specific axioms. Let $Z=\{z_1,\ldots,z_n\}$ be the outcome space (prizes) and $\mathcal{P}$ be the set of probability distributions $F,G,Q:Z\rightarrow[0,1]$ on $Z$ ($F$, $G$ and $Q$ represent the lotteries). We define a binary relation $\succeq$ on $\mathcal{P}$ representing the "preference" of the investor which should satisfy the following axioms
	\\
	\begin{tabularx}{\hsize}{rX|rX}
		1.   &   \textbf{Completeness}
		\begin{equation*}
			F\succeq G \ \text{or} \ G\succeq F
		\end{equation*}     
		&      &   Investor is \textbf{obliged} to choose among $F$ and $G$.\\
		2.   &   \textbf{Transitivity}
		\begin{equation*}
			\text{If} \ F\succeq G \ \text{and} \ G\succeq Q\Rightarrow \ F\succeq Q
		\end{equation*}               
		&      &     \\
		3.   &   \textbf{Continuity}
		
		If $F\succ G\succ Q$, there exist $a,b\in(0,1)$ such that
		\begin{equation*}
			aF + (1-a)Q\succ G\succ bF + (1-b)Q
		\end{equation*}
		&      &   There are no “infinitely good” or “infinitely bad” prizes.                        \\
		4.   &  \textbf{Independence}
		
		Let $a\in(0,1)$. Then,
		\begin{equation*}
			aF + (1-a)Q\succeq aG + (1-a)Q \Leftrightarrow F\succeq G
		\end{equation*}
		&      & If we toss a coin between a fixed lottery $Q$ and lotteries $F$ and $G$ our preference ($F\succeq G$) \textbf{should not} change. (Counterexample: \textbf{Allais paradox})
	\end{tabularx}
	\noindent
	According to VN-M, an investor's preferences will satisfy the above axioms if and only if his overall scope is to maximize his expected utility. In short, the investor will choose among different lotteries using the MEUC or equivalently the VN-M Representation Theorem.
	\begin{theorem}[MEUC]
		A relation $\succeq$ satisfies axioms 1-4 if and only if there exists a utility function $U:Z\rightarrow \mathbb{R}$, such that for every $F,G\in\mathcal{P}$
		\begin{equation*}
			F\succeq G\Leftrightarrow E_F[U(Z)]\geq E_G[U(Z)].
		\end{equation*}
		Moreover, $U$ is unique up to a positive linear transformation, i.e. for some $a>0$ and $b\in\mathbb{R}$
		\begin{equation*}
			\widetilde{U}=aU+b.
		\end{equation*}
	\end{theorem}
	\noindent
	So, the MEUC implies that "an investor with utility function $U$ will prefer lottery $F$ than lottery $G$ if and only if his expected utility for $F$ is larger than that of $G$". The idea that the MEUC is the optimal investment criterion constitutes the cornerstone of economic theory. Based on it, we would like to go a step further and examine the preferences of a class of investors with respect to two lotteries.
	
	For this, we need the notion of Stochastic Dominance (hereafter; SD). More specifically, the SD definition exploits the MEUC in order to derive a conclusion for a specific set of investors $U^*$. With this in mind, we proceed with defining SD.
	\begin{definition}[SD]
		Let two lotteries $Z_1$ and $Z_2$ with cumulative distribution functions $F$ and $G$, respectively. We will say that $Z_1$ dominates $Z_2$, $Z_1DZ_2$, or equivalently, $F$ dominates $G$, $FDG$, if and only if
		\begin{equation*}
			E_F[U(Z_1)]\geq E_G[U(Z_2)],\ \forall U\in U^*
		\end{equation*}
		with a strong inequality for at least one $U_0\in U^*$
	\end{definition}
	The next step is to specify what types of investors could $U^*$ include. In general, we can make the widely accepted and non-restrictive assumption that all investors are wealth maximizers ($U'\geq0$), i.e. they belong to $U^*=\mathbf{U_1}=\{U:U'\geq0\}$. For this set of investors we define the First-order of Stochastic Dominance (hereafter; FSD).
	\begin{definition}[FSD]
		Let two lotteries $Z_1$ and $Z_2$ with cumulative distribution functions $F$ and $G$, respectively. We will say that $Z_1$ first-order stochastically dominates $Z_2$, $Z_1D_1Z_2$, or equivalently, $F$ first-order stochastically dominates $G$, $FD_1G$, if and only if
		\begin{equation*}
			E_F[U(Z_1)]\geq E_G[U(Z_2)],\ \forall U\in \mathbf{U}_1
		\end{equation*}
		with a strong inequality for at least one $U_0\in\mathbf{U}_1$.
	\end{definition}
	The FSD involves the majority of investors, meaning that it applies SD on the largest class of investors possible. From that, we can proceed with narrowing the set of investors. The literature has strong evidence that the majority of investors are also risk-averse ($U''\leq0$), i.e. they belong to $\mathbf{U}_2=\{U:U'\geq 0,U''\leq 0\}$. So, it would be natural to define the Second-order Stochastic Dominance (hereafter; SSD) on $\mathbf{U}_2$.
	\begin{theorem}[SSD]
		Let two lotteries $Z_1$ and $Z_2$ with cumulative distribution functions $F$ and $G$, respectively. We will say that $Z_1$ second-order stochastically dominates $Z_2$, $Z_1D_2Z_2$, or equivalently, $F$ second-order stochastically dominates $G$, $FD_2G$, if and only if
		\begin{equation*}
			E_F[U(Z_1)]\geq E_G[U(Z_2)],\ \forall U\in \mathbf{U}_2
		\end{equation*}
		with a strong inequality for at least one $U_0\in\mathbf{U}_2$.
	\end{theorem}
	Evidently, $\mathbf{U_2}\subset \mathbf{U_1}$ meaning that FSD implies SSD (i.e. $FD_1G\Rightarrow FD_2G$). Now, we can go even further to derive a third-degree of stochastic dominance. But first, we need to give an intuitive description of the new set of investors. As with the assumptions of wealth-maximizing and risk-averse investors, empirical evidence has shown that investors exhibit Decreasing Absolute Risk Aversion (hereafter; DARA). The Absolute Risk Aversion (hereafter; ARA) is defined as shown below
	\begin{equation*}
		r(x)=-\frac{U''(x)}{U'(x)}.
	\end{equation*}
	\noindent
	Since we need $r(x)$ to be decreasing we need
	\begin{equation*}
		r'(x)=\frac{-U'''(x)U'(x)+(U''(x))^2}{(U'(x))^2}<0.
	\end{equation*}
	\noindent
	Based on that $U'\geq 0$ and $U''\leq 0$, this is only possible if $U'''\leq 0$. Consequently, the Third-degree of Stochastic Dominance (hereafter; TSD) can be defined on $\mathbf{U_3}=\{U:U'\geq0,U''\leq0,U'''\geq0\}$.
	\begin{theorem}[TSD]
		Let two lotteries $Z_1$ and $Z_2$ with cumulative distribution functions $F$ and $G$, respectively. We will say that $Z_1$ third-order stochastically dominates $Z_2$, $Z_1D_3Z_2$, or equivalently, $F$ third-order stochastically dominates $G$, $FD_3G$, if and only if
		\begin{equation*}
			E_F[U(Z_1)]\geq E_G[U(Z_2)],\ \forall U\in \mathbf{U}_3
		\end{equation*}
		with a strong inequality for at least one $U_0\in\mathbf{U}_3$.
	\end{theorem}
	\noindent
	This last set of investors represents a very narrow class of investors. Obviously, $\mathbf{U_3}\subset \mathbf{U_2}\subset \mathbf{U_1}$ meaning that FSD implies SSD and both imply TSD (i.e. $FD_1G\Rightarrow FD_2G\Rightarrow FD_3G$). Following the same rationale, we can narrow $U^*$ even further but the aforementioned orders of stochastic dominance are sufficient to discuss how the Markowitz's theory is connected to them.
	
	Based on the Stochastic dominance principle, we could argue that we should stop right here, meaning that every time we need to examine whether or not one lottery stochastically dominates another one we should simply utilize the expected utility of each investor in $U^*$ and see if the same inequality holds. However, if $U^*$ is extremely large (as are the cases of $\mathbf{U_1},\mathbf{U_2}$) it is impossible to check for each and every $U\in U^*$. How do we address this problem? By using the various Stochastic Dominance Rules (hereafter; SDR). The idea is that we can deduce stochastic dominance between two lotteries, $Z_1$ and $Z_2$, for a specific set of investors if and only if the cumulative distributions of lotteries $Z_1$ and $Z_2$ satisfy an appropriate rule. This way we overcome the necessity of testing for each investor in $U^*$ and we focus only on the objective characteristics of the two lotteries under consideration.
	
	When $U^*=\mathbf{U_1}$ we derive the First-order Stochastic Dominance Rule (hereafter; FSDR) as shown below.
	\begin{theorem}[FSDR]
		For any two lotteries $Z_1$ and $Z_2$ with cumulative distributions $F$ and $G$, $F$ dominates $G$ by FSD denoted by $FD_1G$ for all wealth maximizers $U\in\mathbf{U}_1$ ($U'\geq0$) if and only if $F(x)\leq G(x)$ for all values $x$, and there is at least some $x_0$ for which a strong inequality holds. Namely,
		\begin{equation}
			\underset{\forall x, \ \text{with a strong inequality for at least one } x_0}{F(x)\leq G(x)} \ \Leftrightarrow \ \underset{\forall U\in\mathbf{U}_1, \ \text{with a strong inequality for at least one } U_0\in\mathbf{U}_1}{E_{F}[U(Z_1)]\geq E_{G}[U(Z_2)]}
		\end{equation}
	\end{theorem} 
	\noindent
	FSDR says that as soon as the decision maker has an increasing utility function and needs to choose between two lotteries with cumulative distributions $F$ and $G$, he will maximize his expected utility function by choosing the lottery with the smaller cumulative distribution function, i.e. $F$. This degree of stochastic dominance encompasses almost every rational investor, irrespective of his preferences. Two important necessary conditions for FSDR are shown below
	\begin{condition}(Necessary)
		If $FD_1G$, then the expected value of $F$ must be greater than the expected value of $G$. $E_F[x]> E_G[x]$ is a necessary condition for FSD. Equivalently,
		\begin{equation*}
			FD_1G\Rightarrow E_F[Z_1]> E_G[Z_2]
		\end{equation*}
	\end{condition}
	\begin{condition}(Necessary)
		If $FD_1G$, then the the left tail of $G$ must be "thicker". Equivalently,
		\begin{equation*}
			FD_1G\Rightarrow \min_F(x)\geq \min_G(x)
		\end{equation*}
	\end{condition}
	
	Let us turn now to the Second-order Stochastic Dominance Rule (hereafter; SSDR). So, under the additional assumption that investors are risk-averse we derive SSDR as shown below.
	\begin{theorem}[SSDR]
		For any two lotteries $Z_1$ and $Z_2$ with cumulative distributions $F$ and $G$, $F$ dominates $G$ by SSD denoted by $FD_2G$ for all risk-averters $U\in\mathbf{U}_2$ ($U\geq0, U''\leq0$) if and only if
		\begin{equation*}
			\int_{a}^{x}[G(t)-F(t)]dt\geq 0, \quad \forall x\in[a,b],
		\end{equation*}
		with a strict inequality for at least one $x_0$. Equivalently,
		\begin{equation}
			\underset{\forall x, \ \text{with a strong inequality for at least one } x_0}{\int_{a}^{x}[G(t)-F(t)]dt\geq 0} \ \Leftrightarrow \ \underset{\forall U\in\mathbf{U}_2, \ \text{with a strong inequality for at least one } U_0\in\mathbf{U}_2}{E_{F}[U(Z_1)]\geq E_{G}[U(Z_2)]} 
		\end{equation}
	\end{theorem}
	\noindent
	SSDR says that as soon as the decision maker has an increasing and concave utility function and needs to choose between two lotteries with cumulative distributions $F$ and $G$, he will maximize his expected utility function by choosing the lottery with the larger area, i.e. $F$.
	
	As in the case of FSD, we show below the necessary and sufficient conditions for SSDR.
	\begin{condition}(Sufficient)
		If $FD_1G$, then $FD_2G$.
	\end{condition}
	\begin{condition}(Necessary)
		If $FD_2G$, then the expected value of $F$ must be greater than or equal to the expected value of $G$. $E_F[Z_1]\geq E_G[Z_2]$ is a necessary condition for FSD. Equivalently,
		\begin{equation*}
			FD_2G\Rightarrow E_F[Z_1]\geq E_G[Z_2]
		\end{equation*}
	\end{condition}
	\begin{condition}(Necessary)
		If $FD_2G$ and $E_F[Z_1]= E_G[Z_2]$ then the variance of $F$ must be less than or equal to the variance of $G$. $Var_F(Z_1)\leq Var_G(Z_2)$ is a necessary condition for FSD. Equivalently,
		\begin{equation*}
			FD_2G \ \text{and} \ E_F[Z_1]= E_G[Z_2]\Rightarrow Var_F(Z_1)\leq Var_G(Z_2)
		\end{equation*}
	\end{condition}
	\begin{condition}(Necessary)
		If $FD_2G$, then the the left tail of $G$ must be "thicker". Equivalently,
		\begin{equation*}
			FD_2G\Rightarrow \min_F(x)\geq \min_G(x)
		\end{equation*}
	\end{condition}
	\noindent
	So, first of all, FSD implies SSD which is logical as the set of investors that satisfy $U'\geq0$ is a superset of those investors with the additional constraint of $U''\leq0$. Secondly, compared to Condition 2.7, Condition 2.11 does not require a strict inequality and we see from Condition 2.12 that under concave utility functions we derive an extra necessary condition concerning the variances of lotteries.
	
	Last but not least, we can derive the Third-order Stochastic Dominance Rule (hereafter; TSDR) as shown below.
	\begin{theorem}[TSDR]
		For any two lotteries $Z_1$ and $Z_2$ with cumulative distributions $F$ and $G$, $F$ dominates $G$ by TSD denoted by $FD_3G$ for all risk-averters with DARA $U\in\mathbf{U}_3$ ($U\geq0, U''\leq0, U''\geq0$) if and only if
		\begin{align*}
			&(i) \quad \int_{a}^{x}\int_{a}^{z}[G(t)-F(t)]dtdz\geq 0, \ \forall x\\
			&\text{and}\\
			&(ii) \quad E_F[Z_1]\geq E_G[Z_2]
		\end{align*}
		with at least one strong inequality.
	\end{theorem}
	\noindent
	Below we show the necessary and sufficient conditions for TSDR.
	\begin{condition}(Sufficient)
		If $FD_1G$, then $FD_3G$. Also, if $FD_2G$, then $FD_3G$.
	\end{condition}
	\begin{condition}(Necessary)
		If $FD_3G$, then the expected value of $F$ must be greater than or equal to the expected value of $G$. $E_F[x]\geq E_G[x]$ is a necessary condition for FSD. Equivalently,
		\begin{equation*}
			FD_3G\Rightarrow E_F[Z_1]\geq E_G[Z_2]
		\end{equation*}
	\end{condition}
	\begin{condition}(Necessary)
		If $FD_3G$ and $E_F[Z_1]= E_G[Z_2]$ then the variance of $F$ must be less than or equal to the variance of $G$. $Var_F(Z_1)< Var_G(Z_2)$ is a necessary condition for FSD. Equivalently,
		\begin{equation*}
			FD_3G \ \text{and} \ E_F[Z_1]= E_G[Z_2]\Rightarrow Var_F(Z_1)< Var_G(Z_2)
		\end{equation*}
	\end{condition}
	\begin{condition}(Necessary)
		If $FD_3G$ and $E_F[Z_1]= E_G[Z_2]$ and $Var_F(Z_1)= Var_G(Z_2)$ then the skewness of $F$ must be greater than the skewness of $G$. $Skew_F(Z_1)> Skew_G(Z_2)$ is a necessary condition for FSD. Equivalently,
		\begin{equation*}
			FD_3G \ \text{and} \ E_F[Z_1]= E_G[Z_2] \ \text{and} \ Var_F(Z_1)= Var_G(Z_2)\Rightarrow Skew_F(Z_1)> Skew_G(Z_2)
		\end{equation*}
	\end{condition}
	\begin{condition}(Necessary)
		If $FD_3G$, then the the left tail of $G$ must be "thicker". Equivalently,
		\begin{equation*}
			FD_3G\Rightarrow \min_F(x)\geq \min_G(x)
		\end{equation*}
	\end{condition}
	\noindent
	Condition 2.12 does not require equality between variances and Condition 2.18 says that if means and variances remain unchanged then $FD_3G$ necessitates that lottery $Z_1$ is more positively skewed than lottery $Z_2$.
	
	Each of these rules pertains to a specific class of investors. The higher the order of the SD rule the narrower the set of investors. We saw that each of these rules deduces stochastic dominance only by focusing on the objective characteristics of the two lotteries and not on the subjective characteristics of each investor. Moreover, these rules do not specify the types of distributions the two lotteries follow. 
	
	The theory presented in this section can be better interpreted through a concise example. Let the distribution of lotteries $Z_1$, $Z_2$ be defined as in the following table.
	\begin{table}[H]
		\centering
		\caption{Example}
		{
			\begin{tabular}{c c c c}
				\toprule
				\cmidrule{1-4}
				$Z_1$ & $P(Z_1=z_1)$ & $Z_2$ & $P(Z_2=z_2)$\\ 
				$5$ & $0.4$ & $10$ & $0.4$\\
				$10$ & $0.6$ & $20$ & $0.6$\\
			\end{tabular}%
		}
	\end{table}
	\noindent
	So, the means and variances of the two lotteries are, respectively,
	\begin{align*}
		&E[Z_1]=8, \quad Var(Z_1)=6\\
		&E[Z_2]=16 \quad Var(Z_2)=24.
	\end{align*}
	\noindent
	From the above calculations, we observe that $E[Z_2]$ is twice as large as $E[Z_1]$ while at the same time $Var(Z_2)$ is quadruple of $Var(Z_1)$. Can lottery $Z_1$ dominate by any order of stochastic dominance lottery $Z_2$? The answer is no. Because, we saw that under any order of stochastic dominance a necessary condition is $E[Z_1]>E[Z_2]$ (for FOSD) and $E[Z_1]\geq E[Z_2]$ (for SOSD and TOSD). So, it is $Z_2$ that might stochastically dominate $Z_1$. To determine the order of stochastic dominance we need to derive the cumulative distributions of $Z_1$ and $Z_2$. Namely,
	\begin{table}[H]
		\centering
		\caption{Example}
		{
			\begin{tabular}{c c c c}
				\toprule
				\cmidrule{1-4}
				$Z_1$ & $F(z)$ & $Z_2$ & $G(z)$\\ 
				$5$ & $0.4$ & $5$ & $0$\\
				$10$ & $1$ & $10$ & $0.4$\\
				$20$ & $1$ & $20$ & $1$\\
			\end{tabular}%
		}
	\end{table}
	\noindent
	So, we deduce that $G(z)\leq F(z)$, $\forall z$, which means that $Z_2D_1Z_1$. As a result, despite the fact that the variance of lottery $Z_2$ is four times as large as that of lottery $Z_1$, any investor inside $\mathbf{U_3}\subset \mathbf{U_2}\subset \mathbf{U_1}$, regardless of his level of risk-aversion, will prefer $Z_2$.
	
	This last example not only helps in better grasping the concept of SD rules but also indicates that we could probably derive other decision rules that are based on the moments of distributions. A moment-based criterion could simplify even more the decision making of an investor. However, Liu (2004) proved the following Theorem.
	\begin{theorem}
		There is no specific set of moment relationships between the first $n$ moments of lotteries $Z_1$, $Z_2$ with cumulative distribution functions $F$ and $G$ that determines, $FD_1G$, or $FD_2G$, or $FD_3G$.
	\end{theorem}
	\noindent
	In other words, according to Theorem 2.20, we should not expect finding any connection between a moment-based criterion and one of the FSD rules, without making any further assumption with respect to the risk preferences of the investor or the specific type of the distribution of lotteries $Z_1$, $Z_2$. An interesting case that we are going to analyze, extensively, is that of the Mean Variance Rule, introduced by Markowitz (1952, 1959) \cite{Markowitz1952a} and \cite{Markowitz1959}.
	
	\subsection{Mean Variance Rule}
	The notions we discussed in the previous section are so fundamental that any further theoretical finding should be consistent with the MEUC in order for it to be meaningful. In his doctorate thesis in 1952 \cite{Markowitz1952a}, Markowitz proposed a new criterion (rule) with respect to an investor's decision making. This criterion, known as the Mean-Variance Criterion (MVC), suggested that all that the investor needs to know in order to decide between two different lotteries is their first two moments. Particularly, the investor needs to either maximize the expected value of his chosen portfolio (lottery) for a specific level of risk, or equivalently, minimize the risk of the portfolio for a specific level of return. This is formulated as shown below. 
	\begin{definition}[MVC]
		Let $Z_1$ and $Z_2$ be two lotteries with means $\mu_1$, $\mu_2$ and standard deviations $\sigma_1$, $\sigma_2$, respectively. Then, $Z_1$ will satisfy the MV rule (or MVC) with respect to $Z_2$, $Z_1MVZ_2$, if and only if
		\begin{enumerate}\vspace{-5mm}
			\centering
			\item $\mu_1\geq\mu_2$
			\item $\sigma_1\leq\sigma_2$
		\end{enumerate}
	\end{definition}
	
	It is important to detect that through the above definition we do not have any information about the kind of investors that would prefer lottery $Z_1$ instead of lottery $Z_2$. Before discussing the details around this rule we should first highlight why it is so important. The main reason is that if we could specify the class of investors for which this rule is optimal we would only need the first two moments of the two lotteries. So, any further information with respect to the distributions of the two lotteries would simply be irrelevant. 
	
	During Markowitz's doctoral defence he received a rather interesting comment from Milton Friedman, that his contribution was not economics. Potentially, what drove Friedman to make that statement was that Markowitz, at that point, had only made a suggestion that this criterion is meaningful for investors with an expected utility depending only on $\mu$ and $\sigma$, with an increasing and a decreasing relation respectively. A theoretical result was still needed to be found in order to justify the connection between the MVC and the MEUC. In other words, although his idea was intuitive, in the sense that we would expect that (risk-averse) investors desire higher means and detest higher variances, the set of investors for which the MVC coincides with the MEUC was still unknown.
	
	Consider the following example. Assume that we have two risk-averse investors one with $U_1(x)=\ln x$ and the other one with $U_2(x)=\sqrt{x}$. Let also, two cumulative distributions $F$ and $G$ defined as shown in the following table
	\begin{table}[H]
		\centering
		\caption{Example}
		{
			\begin{tabular}{c c c c}
				\toprule
				\cmidrule{1-4}
				$Z_1$ & $F(z_1)$ & $Z_2$ & $G(z_2)$\\ 
				$5$ & $0.80$ & $7$ & $0.99$\\
				$30$ & $0.20$ & $150$ & $0.01$\\
			\end{tabular}%
		}
	\end{table}
	
	From the above Table, we get $E[Z_1]=10>E[Z_2]=8.4$ and $Var(Z_1)=100<Var(Z_2)=203$, and thus MVC is satisfied. Now, the expected utility functions derive
	\begin{align*}
		&E[U_1(Z_1)]=1.9678<E[U_1(Z_2)]=1.9766\\
		&E[U_2(Z_1)]=3.0731>E[U_2(Z_2)]=2.9230.
	\end{align*}
	\noindent
	So, we see that the investor who has a logarithmic utility function will not decide based on the MVC, since if he does so he will select the wrong lottery. On the other hand, the investor with the square root utility function should decide based on the MVC for that specific MV-pair. This example, suffices to conclude that the MVC is not optimal for all risk-averse investors. In the following paragraphs we will discuss which types of investors did Markowitz have in his mind.
	
	The above concerns, drove Markowitz to formulate the MVC under the VN-M theoretical framework in 1959 \cite{Markowitz1959}. He developed his idea by assuming three rational investors each with his own utility function. The first one having a logarithmic utility $\ln{(1+R_p)}$, the second one the square root utility $\sqrt{1+R_p}$ and the third one the cubic root utility $\sqrt[3]{1+R_p}$, where $R_p$ represents the portfolio return. A common characteristic of these three utility functions is that they are all increasing and concave, meaning that Markowitz focused on risk-averse investors. He then applied a Taylor expansion of 2nd-order around $0$, which is defined as
	\begin{equation*}
		U(R_p)\simeq U(0)+U'(0)R_p+\frac{1}{2}U''(0)R_p^2.
	\end{equation*}
	\noindent
	By applying the above Taylor expansion to each one of the three utility functions we obtain
	\begin{alignat*}{3}
		U_1(R_p)&=\ln{(1+R_p)}\\
		&\simeq R_p-\frac{1}{2}R_p^2\\
		U_2(R_p)&=\sqrt{1+R_p}\\
		&\simeq 1+\frac{1}{2}R_p-\frac{1}{8}R_p^2 \\
		U_3(R_p)&=\sqrt[3]{1+R_p}\\
		&\simeq 1+\frac{1}{3}R_p-\frac{1}{9}R_p^2
	\end{alignat*}
	\noindent
	So, each utility function is now expressed as a quadratic approximation. Also, if we apply the expected values on each utility we get
	\begin{alignat*}{3}
		E[U_1(R_p)]&\simeq \mu_p - \frac{1}{2}(\mu_p^2+\sigma_p^2)\\
		E[U_2(R_p)]&\simeq 1 + \frac{1}{2}\mu_p-\frac{1}{8}(\mu_p^2+\sigma_p^2)\\
		E[U_3(R_p)]&\simeq 1 + \frac{1}{3}\mu_p-\frac{1}{9}(\mu_p^2+\sigma_p^2)
	\end{alignat*}
	\noindent
	So, all expected utility functions become a function of only the mean and variance of portfolio returns. Markowitz reasoned that under a no short-selling assumption (i.e. restricting portfolio returns from getting below $-100\%$), for a relatively large range of returns the quadratic approximations are very similar to the respective utility functions. In fact, the following table suggests that.
	\begin{table}[H]
		\centering
		\begin{adjustbox}{scale=0.8}
			\begin{threeparttable}
				\caption{Quadratic Approximations of $\ln{(1+R_p)}$, $\sqrt{1+R_p}$, $\sqrt[3]{1+R_p}$}
				\begin{tabular}{c|c|c|c|c|c|c}
					\toprule
					$R_p$ & $\ln{(1+R_p)}$ & Approx. &  $\sqrt{1+R_p}$ & Approx. &  $\sqrt[3]{1+R_p}$ & Approx.  \\
					\hline
					$-60\%$ & $-.92$ & $-.78$ & $\color{blue}.63$ & $\color{blue}.66$ & $\color{blue}.74$ & $\color{blue}.76$ \\
					$-50\%$ & $-.69$ & $-.63$ & $\color{blue}.71$ & $\color{blue}.72$ & $\color{blue}.79$ & $\color{blue}.81$ \\
					$-40\%$ & $\color{blue}-.51$ & $\color{blue}-.48$ & $\color{blue}.77$ & $\color{blue}.78$ & $\color{blue}.84$ & $\color{blue}.85$ \\
					$-30\%$ & $\color{blue}-.36$ & $\color{blue}-.35$ & $\color{blue}.84$ & $\color{blue}.84$ & $\color{blue}.89$ & $\color{blue}.89$ \\
					$-20\%$ & $\color{blue}-.22$ & $\color{blue}-.22$ & $\color{blue}.89$ & $\color{blue}.90$ & $\color{blue}.93$ & $\color{blue}.93$ \\
					$-10\%$ & $\color{blue}-.11$ & $\color{blue}-.11$ & $\color{blue}.95$ & $\color{blue}.95$ & $\color{blue}.97$ & $\color{blue}.97$ \\
					$0\%$ & $\color{blue}.00$ & $\color{blue}.00$ & $\color{blue}1.00$ & $\color{blue}1.00$ & $\color{blue}1.00$ & $\color{blue}1.00$ \\
					$10\%$ & $\color{blue}.10$ & $\color{blue}.10$ & $\color{blue}1.05$ & $\color{blue}1.05$ & $\color{blue}1.03$ & $\color{blue}1.03$ \\
					$20\%$ & $\color{blue}.18$ & $\color{blue}.18$ & $\color{blue}1.10$ & $\color{blue}1.10$ & $\color{blue}1.06$ & $\color{blue}1.06$ \\
					$30\%$ & $\color{blue}.26$ & $\color{blue}.26$ & $\color{blue}1.14$ & $\color{blue}1.14$ & $\color{blue}1.09$ & $\color{blue}1.09$ \\
					$40\%$ & $\color{blue}.34$ & $\color{blue}.32$ & $\color{blue}1.18$ & $\color{blue}1.18$ & $\color{blue}1.12$ & $\color{blue}1.12$ \\
					$50\%$ & $\color{blue}.41$ & $\color{blue}.38$ & $\color{blue}1.22$ & $\color{blue}1.22$ & $\color{blue}1.14$ & $\color{blue}1.14$\\
					$60\%$ & $.47$ & $.42$ & $\color{blue}1.26$ & $\color{blue}1.26$ & $\color{blue}1.17$ & $\color{blue}1.16$ \\
					$70\%$ & $.53$ & $.46$ & $\color{blue}1.30$ & $\color{blue}1.29$ & $\color{blue}1.19$ & $\color{blue}1.18$ \\
					$80\%$ & $.59$ & $.48$ & $\color{blue}1.34$ & $\color{blue}1.32$ & $\color{blue}1.22$ & $\color{blue}1.20$ \\
					$90\%$ & $.64$ & $.50$ & $\color{blue}1.38$ & $\color{blue}1.35$ & $\color{blue}1.24$ & $\color{blue}1.21$ \\
					$100\%$ & $.69$ & $.50$ & $\color{blue}1.41$ & $\color{blue}1.38$ & $1.26$ & $1.22$ \\
					\hline
				\end{tabular}
				\footnotesize{The \textcolor{blue}{blue} coloured numbers represent "good" approximations, in that they differ from the real value of the utility about $-.03$ to $.03$}
			\end{threeparttable}
		\end{adjustbox}
	\end{table}
	\noindent
	Indeed, the above results indicate the point made by Markowitz. Especially, square and third root utility functions are very well approximated by a quadratic. But why did Markowitz develop his idea this way? The answer to this question comes in the form of the next Theorem proven by Markowitz.
	\begin{theorem}[Markowitz 1959]
		Let $E[R_p]=\mu_p$ and $E[f (R_p)]$, where $R_p$ represents the portfolio returns and $f(R_p)$ a rule which associates a number $f$ to each value of $R_p$. An individual maximizes the expected value of a utility function $U(R_p) = aR_p + bf (R_p)$, if and only if
		\begin{enumerate}[label=(\roman*)]
			\item he maximizes the expected value of some utility function, and
			\item his preferences are based solely on $E[R_p]$ and $E[f(R_p)]$.
		\end{enumerate}
	\end{theorem}
	The above Theorem basically says that an investor with a quadratic utility function that maximizes his utility function will act on the basis of $E[R_p]$ and $E[R_p^2]$. The converse is also true. An investor that bases his preferences solely on $E[R_p]$ and $E[R_p^2]$ and maximizes his expected utility implies that he has a quadratic utility function. So, the appropriate set of investors to which Markowitz was referring to, was the set of quadratic utilities, or as we saw in the previous analysis the set of approximately quadratic utility functions. In the following sections, we will further analyze this subject. 
	
	From this point on, the literature has done an extensive amount of research on Markowitz's framework. It is generally argued that the MVC is meaningful under two alternative hypotheses; (i) the investor's preferences are characterized by a quadratic utility function or, (ii) the distribution of returns is normal. Both of these cases have been proven to be unrealistic and so in general problematic. Starting with the assumption of quadratic preferences we actually defer to increasing absolute risk aversion (hereafter; ARA), which is contrary to the empirical evidence of decreasing ARA. With respect to the Gaussian nature of returns, multiple empirical findings have shown that returns are far from normal, displaying fat-tails, meaning that a Stable-Pareto family of distributions would be more appropriate to assume. 
	
	Now that we have properly set the building blocks that lead to Markowitz's MVC, we can now delve into the fine points of this decision rule. In the following sections, we are going to review and comment on the literature which examines the necessary and sufficient conditions under which the MVC becomes the optimal decision rule. Next, we will carefully test the idea of Markowitz that the MVC under any quadratic approximation of a utility function is almost equivalent to the MEUC.
	
	\subsection{MVC Relation To MEUC: With Known Distribution}
	Having defined the MVC as well as the different orders of SD rules, the question that arises naturally is whether or not there is some kind of connection between these SD rules and the MVC. The importance of this connection stems from the strong theoretical foundation of the SD rules, because if there is not some kind of connection between them and MVC, there is no base in using the MVC.
	
	In 1969, Levy and Hanoch \cite{Hanoch1969} focused on determining when the MVC becomes a necessary and sufficient condition for MEUC. Before them, there were the works of Tobin \cite{Tobin1958} and later Feldstein \cite{feldstein1969}, who concentrated on the type of distribution that makes MVC coincide with MEUC. Tobin, suggested that for any two-parameter distribution MVC coincides with MEUC. The problem with his assertion was an assumption he made, mistakenly, in trying to prove it. That, for any two-parameter distribution function with $\mu$ and $\sigma$ we can standardize, i.e.
	\begin{equation*}
		Z=\frac{X-\mu}{\sigma}.
	\end{equation*}
	\noindent
	But this is not generally true, as for this to hold we need a distribution function for which the location and scale parameters are $\mu$ and $\sigma$, respectively. Some counterexamples of two-parameter distributions are the log-normal, Beta and Gamma. Feldstein noticed that and argued that Tobin's analysis works just for normal distributions. Levy and Hanoch, impacted by Tobin, revolved around the case where the MVC under any two-parameter distribution becomes necessary and sufficient. In fact, they highlighted that the MVC under any two parameter distribution is only a sufficient condition for MEUC. To see that, the following example from the paper of Hanoch and Levy will help.
	\begin{remark}
		Let two random variables $X$ and $Y$ with continuous uniform distributions $F$ and $G$ (i.e. 2-parameter distributions). Let also $X$ have a constant density function in $x_1\leq x\leq x_2$ and $Y$ have a constant density function in $y_1\leq y\leq y_2$, with $x_1>y_1$ and $x_2>y_2$. From this, $F(x)\leq G(x)$ and so $F$ dominates $G$ by FSD. Moreover,
		\begin{equation*}
			\mu_1=\frac{1}{2}(x_1+x_2)>\frac{1}{2}(y_1+y_2)=\mu_2.
		\end{equation*}
		Thus, $\mu_1>\mu_2$ is necessary for dominance. However, the relation between the variances of the two distributions plays no role, meaning that any wealth-maximizer will choose $F$ even if he is highly risk-averse.
	\end{remark}
	\noindent
	In order for MVC to become necessary and sufficient for MEUC, Hanoch and Levy showed that we need to have two 2-parameter distributions with an intersection point. This is formalized in the following Theorem.
	\begin{theorem}
		Let $F$ and $G$ be two distinct distributions with means $\mu_1$, and $\mu_2$,
		and variances $\sigma_1$, $\sigma_2$, respectively, such that $F(x) = G(y)$, for all $x$ and $y$ which satisfy
		$\frac{x-\mu_1}{\sigma_1}=\frac{y-\mu_2}{\sigma_2}$. Let $\mu_1\geq\mu_2$, and $F(x_1)> G(x_1)$ for some $x_1$ (i.e., $F(x)$ and $G(x)$ intersect). Then, $F$ dominates $G$ for all concave $U(x)$, if and only if $\sigma_1\leq\sigma_2$.
	\end{theorem}
	\noindent
	The assumption that "$F(x) = G(y)$, for all $x$ and $y$ which satisfy $\frac{x-\mu_1}{\sigma_1}=\frac{y-\mu_2}{\sigma_2}$" is very restrictive, and at the same time it is hard to interpret. However, without it, it would be impossible to generalize for all two-parameter distributions. 
	
	Levy, knowing that the above Theorem is rather complicated decided that he should restate the Theorem under only normal distributions. In fact, in his book \cite{levy2016} (1998) the new Theorem is structured as shown below.
	\begin{theorem}
		Let two lotteries $Z_1$ and $Z_2$ whose cumulative distributions are $F$ and $G$, respectively, with $Z_1\sim \mathcal{N}(\mu_1, \sigma_1)$ and $Z_2\sim \mathcal{N}(\mu_2, \sigma_2)$. Then, $F$ dominates $G$ by SSD if and only if $F$ dominates $G$ by the MV rule with at least one strong inequality.
	\end{theorem}
	\noindent
	Clearly, the new Theorem avoids any hard to interpret assumptions. More specifically, the nice properties of normal distributions as well as the assumption of "at least one strong inequality" capture the intersection between the two distributions. So, under normality the MVC coincides with the SSDR. Practically this means that any risk averse investor ($U'>0, \ U''<0$), assuming that lotteries are normally distributed should make a decision based either on MVC or SSDR. This explains why the literature insists on the assumption of normality when referring to the MV framework.
	
	However, in the end of Chapter 6 Levy notes: "\textit{actually, the MV coincides with the elliptic family of distributions where the normal distribution belongs to this family}". This statement needs to be investigated thoroughly when it comes to what are the necessary and sufficient conditions for MVC to coincide with MEUC, under any elliptical distribution.
	
	Feldstein argued that one important mistake that Tobin made was that he assumed that any linear combination of random variables following a two-parameter distribution follows the same two-parameter distribution. However, Feldstein pointed that a linear combination of normally distributed random variables remains normal but if we take for example a Gamma distribution any linear combination will have a one-parameter distribution with equal mean and variance. So, Feldstein concluded that the only admissible candidate is a normal distribution. However, Agnew (1971) \cite{Agnew1971} wrote a comment on Feldstein's assertion, claiming that Tobin's Separation Theorem is valid also for non-normal distributions. In particular, he asks the question "If $X_1,\ldots,X_n$, are random variables with finite second moments and if all non-trivial linear combinations $a_1X_1+\ldots+a_nX_n$, have the same distribution except for location and scale, then that distribution must be normal. True or false?". Agnew basically argues that for normality to be the only candidate, the random variables should be stochastically independent. Otherwise, even for uncorrelated random variables the above assertion is false, i.e. there are non-normally distributed random variables that their linear combinations follow the same distribution. A specific example is the standardized bilateral exponential distribution (or else Laplace distribution with $\mu=1$ and $\beta=1$), which belongs to the elliptic family of distributions.
	
	Later, Chamberlain in 1983 \cite{Chamberlain1983} (also Owen-Rabinovicth \cite{Owen_Rab_1983}) introduced two Theorems, regarding the relation between elliptical distributions and the MVC, giving substance to Agnew's assertion. But first we need to define the spherical and elliptical distributions.
	\begin{definition}[Spherical distributions]
		A random vector $X$ is \textbf{spherically distributed} about the origin if its probability density function $f$ satisfies the following
		\begin{equation*}
			f(X)=f(MX),
		\end{equation*}
		where $M^\top M=MM^\top=I_n$.
	\end{definition}
	\noindent
	Equivalently, a spherical distribution is invariant under orthogonal linear transformations that leave the origin fixed. Likewise, an elliptical distribution is defined as shown below.
	\begin{definition}[Elliptical distributions]
		A random vector $X(n\times 1)$ is \textbf{elliptically distributed} if
		\begin{equation*}
			X=\mu+AY,
		\end{equation*}
		where $Y$$(k\times 1)$ is a spherically distributed random vector, $A$ is a $(n\times k)$ matrix such that $AA^\top=\Sigma$ (with $\Sigma$ representing the scale matrix), and $\mu$$(n\times 1)$ is the location vector.
	\end{definition}
	\begin{remark}
		\begin{itemize}
			\item All symmetric elliptical distributions are \textit{symmetric} around $\mu$. So,
			\begin{equation*}
				E[(X-\mu)^i]=0, \ \text{for} \ i=3,5,7\ldots
			\end{equation*}
			\item All symmetric elliptical distributions are determined exactly by their mean and variance
			\item Any linear combination of elliptically distributed variables is still elliptical
			\item Under elliptical distributions, variance constitutes the precise measure of risk.
			\item Some elliptical distributions are: Normal, Student's t, Laplace, Logistic, Exponential, etc.
		\end{itemize}
	\end{remark}
	
	Now, the first Theorem of Chamberlain, considers the case where portfolio returns are made up of risky assets and a risk-free asset. Namely, 
	\begin{theorem}[MV-utilities under elliptical distributions]
		The distribution of portfolio returns $R_p=w'R+(1-w)R_f$ is determined by its
		mean $\mu_p$ and variance $\sigma_p$ for every $w$ if and only if there is a non-singular matrix $T$ such that
		\begin{equation*}
			z=T(R-\mu),
		\end{equation*}
		is spherically distributed about the origin.
	\end{theorem}
	\noindent
	We saw that a linear transformation of a spherical random vector is elliptically distributed, which means that the asset returns $R$ are elliptically distributed and since any linear combination of elliptical distribution is also elliptical that also makes $R_p$ being elliptically distributed. Thus, the above Theorem states that if there is a riskless asset in the investor's portfolio and the distribution of the risky assets is elliptical, the distribution of the portfolio's returns will be determined only by $\mu_p$ and $\sigma_p$. Accordingly, that derives the following result 
	\begin{equation*}
		E[U(R_p)]=f(\mu_p,\sigma_p).
	\end{equation*}
	Does that implicate that the MVC coincides with the MEUC? The answer to this question is no. In order to accept that, we should prove that under elliptical distributions the following equivalence holds
	\begin{equation*}
		Z_1MVZ_2 \Leftrightarrow FD_2G.
	\end{equation*}
	Fortunately, the "necessity" side has been proven by Chamberlain, namely $Z_1MVZ_2 \Rightarrow FD_2G$. In particular, Chamberlain showed that for any concave utility function, i.e. $U\in\mathbf{U_2}$, the expected utility is increasing in mean and decreasing in variance. In other words, the MVC implicates the MEUC. But what about the "sufficiency" side, namely $FD_2G \Rightarrow Z_1MVZ_2$? For this, we will need to make use of Conditions 2.11 and 2.13. Condition 2.13, known as the left-tail necessary condition for the SSD, entails that $\sigma_1\leq\sigma_2$, since we are talking about elliptical distribution which are known to be determined by their mean and variance. Accordingly, Condition 2.11 entails that $\mu_1\geq\mu_2$. Thus, we get also the "sufficiency" side. And so now, we can claim that under elliptical distributions
	\begin{equation*}
		Z_1MVZ_2 \Leftrightarrow FD_2G.
	\end{equation*}
	So, for any two lotteries which are elliptically distributed, the optimal rule for a risk-averse investor is the MVC. In other words, the Theorem 2.25 can be restated as shown below.
	\begin{theorem}[MVC-SSD under elliptical distributions]
		Let two lotteries $Z_1$ and $Z_2$ whose cumulative distributions are denoted by $F$ and $G$, respectively. Let also $Z_1,Z_2$ be elliptically distributed with $\mu_1, \sigma_1$ and $\mu_2, \sigma_2$, respectively. Then, $F$ dominates $G$ by SSD if and only if $F$ dominates $G$ by the MV rule with at least one strong inequality.
	\end{theorem}
	The above Theorem states that, regardless of the distribution being normal, or logistic, or Laplace, or any other type of elliptical distribution, the investor should use the MVC to make his decisions. This elevates the value of the mean-variance method developed by Markowitz. However, in practice, we can see in the following Remark that this family of distributions is quite limited since in order for the MVC to be meaningful we need skewness to be equal to zero.
	\begin{remark}
		Some valid cases of symmetric elliptical distributions are:
		\begin{itemize}
			\item Student's-t: If $df> 3$, then $\mu=0$ $\sigma^2=\frac{df}{df-2}$, $s=0$, $\kappa=3+\frac{6}{df-4}$, if $df>4$, otherwise undefined
			\item Laplace: $\mu=\mu$, $\sigma^2=2b^2$, $s=0$, $\kappa=6$, $b>0$
			\item Logistic: $\mu=\mu$, $\sigma^2=\frac{a^2\pi^2}{3}$, $s=0$, $\kappa=\frac{21}{5}$, $a>0$
			\item $\alpha$-stable: $\mu=\mu$, $\sigma^2=2c^2$, $s=0$, $\kappa = 3$, if $\alpha=2$ (Gaussian case)
		\end{itemize}
	\end{remark}
	Duchin and Levy (2004) \cite{Levy2004a} conducted an empirical study to determine how this new finding from Chamberlain correlates with real data. They used monthly returns for 5 portfolios spanning from 1926 to 2001. Namely, common stocks, small stocks, long-term corporate bonds, long-term government bonds and Treasury bills. Then, they tested which of the following candidate distributions: Normal, Beta, Exponential, Extreme value, Gamma, Logistic, Lognormal, Student-t, Skew-Normal, Stable Paretian and Weibull, best fits the data. They found strong evidence pointing to the logistic distribution which belongs to the symmetric elliptical family of distributions. Based on Theorem 2.30, they argued that this indicates that the MVC is the optimal decision rule for these portfolios.
	
	Although it is clear that the MVC is optimal under elliptical distributions, many research papers and academic books still consider the MV-framework only under either quadratic preferences or normality. Markowitz (2010) \cite{Markowitz2010} has observed that and that is why he emphasizes that "I never-at any time!-assumed that return distributions are Gaussian". True, the literature has often misinterpreted under what conditions the MVC is valid, but even under elliptical distributions the MVC is still far from being truly useful when dealing with real stock or portfolio returns. More specifically, families of distributions which contain more non-normal cases are more interesting, since they are known to describe better empirical data.
	
	One very recent work from Schuhmacher et al (2021) \cite{Schuhmacher2021} tries to broaden the family of distributions for which the MVC is relevant. The authors show that, in the presence of a risk-free asset, the return distribution of every portfolio is determined by its mean and variance if and only if asset returns follow a specific Skew-Elliptical distribution. A Skew-Elliptical distribution is defined as shown below 
	\begin{definition}[Skew-Elliptical GLS distributions]
		A random vector $X(n\times1)$ is said to have a Skew-Elliptical generalized location-scale (hereafter; GLS) distribution with constant $r\in\mathbb{R}$, if its components $X_i$ $(i=1,\ldots,n)$, can be written as
		\begin{equation*}
			X_i=r+\beta_iY+\gamma_iZ_i,
		\end{equation*}
		where, conditional on $Y$,the vector $Z=(Z_1, \ldots,Z_n)'$ is spherically distributed and $Y$ is a real-valued random variable with $E[Y]\neq0$ and $Var[Y]=1$. The coefficients $\beta_i$, $gamma_i$ are real numbers with $\beta_i\neq 0$ for at least one $i=1,\ldots,n$.
	\end{definition}
	\begin{remark}
		\begin{itemize}
			\item All Skew-Elliptical distributions are determined exactly by their mean and variance. More specifically, based on Definition 2.32 we derive
			\begin{align*}
				&E[X_i] = r + \beta_iE[Y] + \gamma_iE[Z_i]\\
				&Var(X_i) = \beta_i^2Var(Y) + \gamma_i^2Var(Z_i)=\beta_i^2+\gamma_i^2.
			\end{align*}
			Solving for $\beta_i$ and $\gamma_i$ we get
			\begin{align*}
				&\beta_i = \frac{E[X_i]-r}{E[Y]}\\
				&|\gamma_i| = \sqrt{Var(X_i)-\Big(\frac{E[X_i]-r}{E[Y]}\Big)^2}.
			\end{align*}
			\item Any linear combination of Skew-elliptically GLS distributed variables is still Skew-Elliptical
			\item Some Skew-Elliptical distributions are: Skew-Normal, Skew-t, Skew-Cauchy, Skew-logistic, etc.
		\end{itemize}
	\end{remark}
	Similar to Chamberlain, Schuhmacher et al. proved the following Theorem.
	\begin{theorem}[MV-utilities under Skew-Elliptical distributions]
		Assume there exists at least one $i=1,\ldots,n$ such that $E[R_i]\neq R_f$, where $R_i$ is the $i$th element of the risky asset vector $R$. In the presence of a risk-free asset, $R_f$, the distribution of portfolio returns  $R_p=w'R + (1-w)R_f$ is determined by its mean and variance for every $w\in \mathbb{R}^{n+1}$ with $w'\mathds{1}=1$ if and only if the asset returns $R$ have a Skew-Elliptical GLS distribution.
	\end{theorem}
	Theorem 2.34 states that lotteries which follow a Skew-Elliptical distribution have a MV-utility. So, we derive the following result
	\begin{equation*}
		E[U(R_p)]=f(\mu_p,\sigma_p).
	\end{equation*}
	\noindent
	Following the same rationale as in Chamberlain's work, the fact that the expected utility is only a function of the mean and the variance of the portfolio returns does not implicate that the MVC is necessary and sufficient for the MEUC. In other words, one needs to prove that under skew-elliptical distributions the following holds
	\begin{equation*}
		Z_1MVZ_2 \Leftrightarrow FD_2G.
	\end{equation*}
	Contrary to Chamberlain, Schuhmacher et al. do not show, that $f$ is increasing in mean and decreasing in variance, for any $U\in\mathbf{U_2}$. So, with regards to the "sufficiency" side, this might mean that $FD_2G$ does not necessarily implicate $Z_1MVZ_2$. Moreover, when it comes to proving the "necessity" side we cannot make use of Condition 2.13, as we did earlier for the elliptical family of distributions, since this Condition only applies to distributions that are not skewed. In fact, later on we will see through Monte Carlo Simulations that under Skew-Elliptical distributions the "necessity" side is violated. So, contrary to elliptical distributions, we cannot prove that under Skew-Elliptical distributions the MVC is the optimal decision rule for any risk-averse investor. We only know that under Skew-Elliptical distributions the expected utility of the investor is a function of mean and variance. As a result, there might be cases in which even though the MVC is satisfied between two lotteries, namely $Z_1MVZ_2$, some type of investor inside $\mathbf{U_2}$ might prefer lottery $Z_2$. This will become evident in our Monte Carlo simulations in the Quadratic approximations subsection.

	\subsection{MVC Relation To MEUC: With Known Preferences}
	An alternative to searching for a good candidate distribution is to make an assumption on the utility function of the investor. A widely used premise is that of quadratic utility. In that case, the expected utility becomes a function of only $\mu$ and $\sigma$. The price we pay for this kind of assumption is that, (i) quadratic preferences constitute a very restrictive class and, (ii) by assuming quadratic utility we are led to increasing absolute risk aversion (ARA), which is counter-intuitive. So, one should be careful when trying to avoid an assumption with respect to the distribution of returns, as he will be left with a class of utility functions that is questionable for its realism as well as for its usefulness. However, this is the specific class of investors that Markowitz pointed to.
	
	Hanoch and Levy (1969) \cite{Hanoch1969}, presented an example through which they argued that the MVC under quadratic preferences is only sufficient for MEUC. This, can be formally shown through the following Proposition from Hanoch and Levy (1970) \cite{Hanoch_Levy1970}.
	\begin{proposition}
		Assuming quadratic preferences, the MVC is only a sufficient condition for the MEUC
	\end{proposition}
	\begin{proof}
		(Sufficiency) Following \cite{Hanoch_Levy1970}, let the following quadratic utility function
		\begin{equation*}
			U(x)=2Kx-x^2
		\end{equation*}
		where $K>0$, $U'(x)=2(K-x)>0$ and $U''(x)=-2<0$. Let two portfolios $x_1$ and $x_2$ for which we derive
		\begin{align*}
			\Delta E[U]&=E[U(x_1)]-E[U(x_2)]\\
			&=2K\mu_1-E[x_1^2]-2K\mu_2-E[x_2^2]\\
			&=2K\mu_1-(\mu_1^2+\sigma_1^2)-2K\mu_2-(\mu_2^2+\sigma_2^2)\\
			&=2K\Delta\mu-(\Delta\mu^2+\Delta\sigma^2)\\
			&=2\Delta\mu(K-\bar{\mu})-\Delta\sigma^2
		\end{align*}
		where $\Delta\sigma^2=\sigma_1^2-\sigma_2^2$, $\bar{\mu}=\frac{\mu_1+\mu_2}{2}$. Since $\mu_1,\mu_2<K$, we have that $\bar{\mu}<K$. Then, $\Delta E[U]>0$ if we assume that $\mu_1>\mu_2$ and $\sigma_1<\sigma_2$, which is exactly the MVC.
		
		(Necessity) Let $\Delta E[U]>0$. Does that imply $\mu_1>\mu_2$ and $\sigma_1<\sigma_2$? The answer is no. In fact, from $\Delta E[U]>0$ we have that 
		\begin{align*}
			&2\Delta\mu(K-\bar{\mu})>\Delta\sigma^2,\\
			&\Delta\mu>0
		\end{align*}
		Thus, even if $\Delta\sigma^2<0$ (i.e. $\sigma_1>\sigma_2$), since $2\Delta\mu(K-\bar{\mu})>0$, the above inequality holds.
	\end{proof}
	The above proof led Hanoch and Levy to identify the right rule which is both necessary and sufficient for MEUC. The rule is called quadratic dominance rule and we can see below how it coincides with MEUC.
	\begin{theorem}
		Assuming quadratic preferences, the quadratic dominance rule as defined below
		\begin{align*}
			&\textit{1.}\ \mu_1\geq\mu_2\\
			&\textit{2.}\  2\Delta\mu\big(\max(x_1,x_2)-\bar{\mu}\big)-\Delta\sigma^2\geq0,
		\end{align*}
		is both necessary and sufficient for MEUC.
	\end{theorem}
	The proof of this Theorem is evident from the previous proof of the Proposition. The reason the authors chose to replace $K$ with $\max(x_1,x_2)$ is that in this way the rule constitutes a smaller set than if we had $K$. This set happens to be the smallest and thus the optimal set. The new rule under quadratic preferences is both necessary and sufficient for MEUC, mainly because it also includes cases where $\sigma_1>\sigma_2$. 
	
	Johnstone et al. (2011) \cite{Johnstone2011} proved that if one wants to avoid constraining the distribution of portfolio returns it is necessary to assume quadratic preferences to apply the MVC. This is formulated as shown below.
	\begin{theorem}
		The use of MVC, on the class of all distributions, implies that the decision maker’s utility function must be quadratic.
	\end{theorem}
	\noindent
	This Theorem basically says that we should not look for any other set of utilities other than the quadratic, assuming that we do not constrain the family of distributions, which justifies the use of quadratic utilities under the MV-framework. Still, the quadratic family of utilities is very restrictive. But in Markowitz's words (2010) \cite{Markowitz2010}: "Nor did I ever assume that the investor’s utility function is quadratic". So, although the literature has adopted the quadratic utility as the only appropriate class of investors for which the MV-framework is relevant, Markowitz claims that we should not be fixated just on the quadratic utility function. In particular, as we have already discussed, Markowitz attempted to upgrade the MVC by discussing its validity even under a wider class of utility functions that happen to be approximately quadratic. We will discuss this premise in the following section.
	
	\section{Empirical Results And Methodology}
	\subsection{Approximately Quadratic Utility Functions}
	As we analyzed earlier, Markowitz (1959) chose three specific utilities in his work in order to discuss his idea about the quadratic approximations, namely $\log(1+Z)$, $\sqrt{1+Z}$ and $\sqrt[3]{1+Z}$. Those three utilities are not only concave but they also satisfy one additional property, namely $U'''\geq0$. To see why this extra property is crucial, we have to take a 2nd-order Taylor series on the utility function, as shown below
	\begin{align*}
		&Q_Z=U(\mu)+U'(\mu)(Z-\mu)+\frac{U''(\mu)}{2}(Z-\mu)^2\\
		&E[Q_Z]=U(\mu)+\frac{U''(\mu)}{2}\sigma^2
	\end{align*}
	\noindent
	So, $E[Q_Z]$ will increase with respect to $\mu$ and decrease with respect to $\sigma$ if the following holds
	\begin{align*}
		&\frac{\partial E[Q_Z]}{\partial \sigma}=U''(\mu)\sigma<0, \ \text{if} \ U''<0\\
		&\frac{\partial E[Q_Z]}{\partial \mu}=U'(\mu)+\frac{U'''(\mu)}{2}\sigma^2>0, \ \text{if} \ U'>0 \ \text{and} \ U'''\geq 0
	\end{align*}
	Therefore, any utility function that is a part of $\mathbf{U_3}=\{U:U'>0,U''<0,U'''\geq 0\}$ and at the same time is almost quadratic, will be increasing in mean and decreasing in variance. Are $\log(1+Z)$, $\sqrt{1+Z}$ and $\sqrt[3]{1+Z}$ almost quadratic? According to Markowitz's Table 4, we see that $\sqrt{1+Z}$ and $\sqrt[3]{1+Z}$ are approximately quadratic for any value around $-60\%$ and $100\%$, but with regards to $\log(1+Z)$ the quadratic approximation is good only for values around $-40\%$ and $50\%$. As a result, assuming that two lotteries $Z_1$, $Z_2$ take values in $[-60\%,100\%]$ (or in $[-40\%,50\%]$) and that $Z_1MVZ_2$, we could say that the investors with utility functions $\sqrt{1+Z}$ and $\sqrt[3]{1+Z}$ (or $\log(1+Z)$) should prefer $Z_1$. Does the inverse also hold? Based on Proposition 2.35, we deduce that an approximately quadratic utility function can qualify the MVC to be only a sufficient condition for the MEUC. So, according to Markowitz, without any further assumption on the distributions followed by $Z_1$ and $Z_2$, the MVC will be sufficient for the MEUC if and only if the utility function of the investor is approximately quadratic for a sufficiently wide range of values. In the end of this work there is an Appendix which includes several plots of different utility functions together with their quadratic approximation. Markowitz realized that this reasoning does not suffice to support his assertion. Thus, he resorted to a quite different approach.
	
	In 1979, Levy and Markowitz \cite{Levy1979} revisited this subject by doing an empirical analysis. Firstly, they restated Markowitz's premise as follows: "an investor that chooses carefully from among the mean-variance efficient set, will almost maximize his expected utility, if and only if his utility function is approximately quadratic", i.e. it is almost perfectly approximated by a 2nd-order Taylor expansion. Consequently, they introduced a way to identify the size of the set containing the approximately quadratic utility functions. Firstly, the utility functions have to be a part of $\mathbf{U_3}$. Such utility functions are the following, $\log{(1+Z)}$, $(1+Z)^a$ with $a=0.1,0.3,0.5,0.7,0.9$ and $-e^{-a(1+Z)}$ with $a=0.1,0.5,1,3,5,10$. Secondly, they collected the annual returns of $149$ mutual funds during the period 1958 through 1967. Following that, they calculated $Corr(E[U(Z)],E[Q_Z])$, for each of the above utility functions. The idea was simple. If the correlation of the expected utility and the expected value of the quadratic approximation is close to $1$, that would indicate that $E[U(Z)]$ and $E[Q_Z]$ move in the same direction, which is the actual point of interest. This approach, overcomes the limitations in Markowitz's initial attempt to promote his idea of approximately quadratic utilities. Going back to their results, the authors found evidence of
	\begin{equation*}
		Corr(E[U(Z)],E[Q_Z])\simeq 1,
	\end{equation*}
	for all the parametrizations of the utility functions, except for $b=5,10$ which represent the extremely risk-averse investors. Additional empirical evidence came from their joint work with Kroll in 1984 \cite{Kroll_1984}. Based on these findings, they argued that the above utility functions are almost quadratic for almost all of their parametrizations. Thus, if the MVC holds, the above investors should decide based on it.
	
	In the following section we are going to thoroughly analyze our approach on this subject. Up to this point, the context of our discussion around the connection between the MVC and MEUC includes either an assumption with respect to the set of utility functions or the type of distribution of the lotteries. With that being said, although Levy and Markowitz did provide some supportive evidence of their premise, we believe that in order for it to be confirmed we need to clarify whether we need an extra assumption with respect to the type of distribution. Namely, since we are assuming approximately quadratic utility functions we are obliged to research on whether or not we need an extra assumption on the kind of distribution under which the MVC is sufficient for the MEUC. Otherwise, the premise about quadratic approximations cannot be strongly supported. For our analysis, we need to define $\mathbf{U^*_3}=\{U:U'>0,U''<0,U'''\geq 0 \ \text{and} \ U(Z)\simeq Q_Z\}$ to be the set that contains all those utility function that are part of $\mathbf{U_3}$ and at the same time are almost quadratic.
	
	\subsection{Methodology}
	An important question that needs to be answered is under what conditions does the assumption of approximately quadratic utility functions hold. Whether or not a utility function is well-approximated by a 2nd-order Taylor series should depend on the utility function we use but also the type of the assumed distribution. A simple approach like that on Table 4 is inadequate. The reason is that non-normal or skewed distributions might conflict with quadratic approximations, in terms of the validity of the MVC. In other words, we should specify for which distributions we have $U(Z)\simeq Q_Z$. In a recent review of his work, Markowitz (2010) \cite{Markowitz2010}, claimed that the idea presented in \cite{Levy1979} was targeting any type of distribution. This last information helps us formulate a mathematical Proposition connecting the MVC to approximately quadratic utility functions. Before doing that we need to highlight the following. First, we should take into consideration the fact that the MVC is a decision criterion between two lotteries. So, the premise of Markowitz should be restated accordingly. Second, based on Proposition 2.35, the MVC is only a sufficient condition for MEUC, for any quadratic utility function. Third, the set of investors we refer to is $\mathbf{U_3^*}=\{U:U\in\mathbf{U_3} \ \text{and} \ U(Z)\simeq Q_Z\}$. Altogether, we get the following Corollary.
	\begin{corollary}[MVC under Quadratic Approximation]
		For any two lotteries $Z_1$ and $Z_2$, with any cumulative distributions $F$ and $G$, the following holds
		\begin{equation*}
			Z_1MVZ_2 \Rightarrow E_F[U(Z_1)]\geq E_G[U(Z_2)], \ \forall U\in\mathbf{U_3^*}=\{U:U\in\mathbf{U_3} \ \text{and} \ U(Z)\simeq Q_Z\}.
		\end{equation*}
	\end{corollary}
	\begin{proof}
		If $Z_1MVZ_2$ then for any $U\in \mathbf{U_3^*}$ we have
		\begin{align*}
			E_F[U(Z_1)]-E_G[U(Z_2)]&\simeq E_F[Q_{Z_1}]-E_G[Q_{Z_2}]\\
			&=U(\mu_1)+\frac{U''(\mu_1)}{2}\sigma_1^2-U(\mu_2)-\frac{U''(\mu_2)}{2}\sigma_2^2\geq 0
		\end{align*}
	\end{proof}
	\noindent
	The above Corollary states that for any two lotteries that satisfy the MVC, i.e. $Z_1MVZ_2$, all investors inside $\mathbf{U_3^*}$ will maximize their expected utility functions by choosing lottery $Z_1$. But the success of Corollary 3.1 relies on $\mathbf{U_3^*}$ being sufficiently large. This is what we will try to determine. We already know from Theorem 2.30 that for any elliptical symmetric distribution the MVC becomes equivalent to the MEUC. So, for these types of distributions the additional limiting assumption of approximately quadratic utility functions is unnecessary. In other words, we would like to examine the validity of the above Corollary for asymmetric distributions and even for very non-normal cases which are considered to characterize daily or even monthly stock returns. Jondeau and Rockinger (2006) \cite{Jondeau2006}, using empirical data, supported that cubic or even quartic approximations are better approximations of expected utility, under large departure from normality. But their work does not approach the work of Levy and Markowitz the way we do. As long as the quadratic approximation consistently results in the same decision making between two lotteries as the direct MEUC, there is no reason in searching for more precise approximations of the utility function. The most efficient way to research that is by applying Monte Carlo simulations. The simulations enable us to apply different types of distribution with specific characteristics. This way we can identify more clearly under what conditions the premise of Markowitz is valid.
	
	\subsection{Empirical Evidence}
	We will proceed with presenting some empirical findings that motivate our thinking. We have collected monthly stock returns spanning from 2000 to 2021 for 850 US stocks that were (or still are) constituents of NYSE. Stocks are sorted with respect to their individual skewnesses. Then, we create deciles, with the first one containing the more negatively-skewed stocks and the last decile containing the more positively-skewed stocks. The statistics for each decile are shown in the following table.
	\begin{table}[H]
		\centering
		\scalebox{0.9}{
			\begin{threeparttable}
				\caption{NYSE constituents average statistics per decile}
				\begin{tabular}{ c c c c c c c c c c c}
					\toprule
					& Dec 1 & Dec 2 & Dec 3 & Dec 4 & Dec 5 & Dec 6 & Dec 7 & Dec 8 & Dec 9 & Dec 10 \\
					\hline
					Mean & 0.0083 & 0.0101 & 0.0101 & 0.0099 & 0.0098 & 0.0114 & 0.0120 & 0.0120 & 0.0156 & 0.0163 \\
					Std & 0.0807 & 0.0847 & 0.0890 & 0.0895 & 0.0981 & 0.1071 & 0.1095 & 0.1274 & 0.1448 & 0.1772 \\
					Skewness & -0.7251 & -0.3041 & -0.1502 & -0.0062 & 0.1013 & 0.2223 & 0.3849 & 0.6205 & 1.096 & 2.8186 \\
					\hline
				\end{tabular}
		\end{threeparttable}}
	\end{table}
	
	By doing that, we imply that maybe skewness should play an important role in the decision making of the investors inside $\mathbf{U_3}$. In fact, any utility function in $\mathbf{U_3}$ is increasing in skewness. So, we are interested into the cases in which the asset from the first decile satisfies the MV rule with respect to the asset on one of the other deciles. This way, we can detect whether the MVC is sufficient for the MEUC or if the higher skewness is more desirable by the investors inside $\mathbf{U_3}$, leading to the failure of Corollary 3.1. Schematically, we look for the following MV-pairs
	\begin{align*}
		&\text{\underline{Decile 1}} & \text{\underline{Deciles 1 \& 2}} && \ldots && \text{\underline{Deciles 1 \& 10}}\\
		&Z^{dec_1}MVZ^{dec_1} & Z^{dec_1}MVZ^{dec_2} && \ldots && Z^{dec_1}MVZ^{dec_{10}}
	\end{align*}
	The utility functions we are going to use are a combination of the ones used by Levy and Markowitz (1979) and Ederington (1995). The utilities are presented in the following table
	\begin{table}[H]
		\centering
		\scalebox{0.85}{
			\begin{threeparttable}
				\caption{Utility functions inside $\mathbf{U_3}$}
				\begin{tabular}{c c}
					\toprule
					$(1+Z)^a$ with $a=\{0.01, 0.1, 0.5, 0.9\}$ &\\
					$\log(a+Z)$ with $a=\{0.9, 1\}$ &\\
					$-e^{-a(1+Z)}$ with $a=\{0.7, 1, 3, 5, 8, 10, 15, 20\}$ &\\
					$-(1+Z)^{-a}$ with $a=\{0.01, 0.3, 0.5, 1, 3, 5, 8, 10, 15, 20\}$ &\\
					\bottomrule
				\end{tabular}
		\end{threeparttable}}
	\end{table}
	Before going into the empirical results we have to highlight a few things about the level of risk-aversion of each utility function. We can measure the level of risk-aversion of each utility function by the absolute risk aversion. Namely,
	\begin{align*}
		&ARA_{(1+Z)} = \frac{1-a}{1+Z}\\
		&ARA_{log} = \frac{1}{a+Z}\\
		&ARA_{exp} = a\\
		&ARA_{-(1+Z)} = \frac{1+a}{1+Z}\\
	\end{align*}
	\noindent
	In our case, lotteries $Z$ represent stock returns which means that the range of values is $[-1,1]$. So, in general we can sort the utility functions in terms of their level of risk-aversion as follows $(1+Z)^a$, $\log(a+Z)$, $-e^{-a(1+Z)}$ and $-(1+Z)^{-a}$, with the last one describing the more risk risk-averse investor. Parameter-wise the $log$-utility function characterizes the more risk-averse investors when $a$ gets closer to $0.9$. Accordingly, the $(1+Z)^a$ is more risk-averse for $a$'s closer to $0.01$. For $-e^{-a(1+Z)}$ and $-(1+Z)^{-a}$ the higher $a$ gets, the more risk-averse the investors are. One would expect that for more risk-averse investors the skewness of an asymmetric distribution together with the existence of more extreme jumps will impact their decision making.
	
	In the following tables, we can see the percentage of times that the MVC deduces the MEUC, for each utility function per decile for the multiple MV-pairs we get from our data.
	
	\begin{table}[H]
		\centering
		\scalebox{0.9}{
			\begin{threeparttable}
				\caption{Percentage of MVC$\Rightarrow$MEUC for $(1+Z)^a$}
				\begin{tabular}{ c c c c c}
					\toprule
					& $a=0.01$ & $a=0.1$ & $a=0.5$ & $a=0.9$\\
					\hline
					Dec 1 vs Dec 1 & 100\% & 100\% & 100\% & 100\% \\
					Dec 1 vs Dec 2 & 100\% & 100\% & 100\% & 100\% \\
					Dec 1 vs Dec 3 & 100\% & 100\% & 100\% & 100\% \\
					Dec 1 vs Dec 4 & 100\% & 100\% & 100\% & 100\% \\
					Dec 1 vs Dec 5 & 100\% & 100\% & 100\% & 100\% \\
					Dec 1 vs Dec 6 & 100\% & 100\% & 100\% & 100\% \\
					Dec 1 vs Dec 7 & 100\% & 100\% & 100\% & 100\% \\
					Dec 1 vs Dec 8 & 99\% & 100\% & 100\% & 100\% \\
					Dec 1 vs Dec 9 & 100\% & 100\% & 100\% & 100\% \\
					Dec 1 vs Dec 10 & 100\% & 100\% & 100\% & 100\% \\
					\hline
				\end{tabular}
		\end{threeparttable}}
	\end{table}
	
	\begin{table}[H]
		\centering
		\scalebox{0.9}{
			\begin{threeparttable}
				\caption{Percentage of MVC$\Rightarrow$MEUC for $log(a+Z)$}
				\begin{tabular}{ c c c}
					\toprule
					& $a=0.9$ & $a=1$ \\
					\hline
					Dec 1 vs Dec 1 & 100\% & 100\% \\
					Dec 1 vs Dec 2 & 100\% & 100\% \\
					Dec 1 vs Dec 3 & 100\% & 100\% \\
					Dec 1 vs Dec 4 & 100\% & 100\% \\
					Dec 1 vs Dec 5 & 100\% & 100\% \\
					Dec 1 vs Dec 6 & 99\% & 100\% \\
					Dec 1 vs Dec 7 & 100\% & 100\% \\
					Dec 1 vs Dec 8 & 99\% & 100\% \\
					Dec 1 vs Dec 9 & 100\% & 100\% \\
					Dec 1 vs Dec 10 & 99\% & 100\% \\
					\hline
				\end{tabular}
		\end{threeparttable}}
	\end{table}

	\begin{table}[H]
		\centering
		\scalebox{0.9}{
			\begin{threeparttable}
				\caption{Percentage of MVC$\Rightarrow$MEUC for $-e^{-a(1+Z)}$}
				\begin{tabular}{ c c c c c c c c c}
					\toprule
					& $a=0.7$ & $a=1$ & $a=3$ & $a=5$ & $a=8$ & $a=10$ & $a=15$ & $a=20$ \\
					\hline
					Dec 1 vs Dec 1 & 100\% & 100\% & 100\% & 99\% & 97\% & 95\% & 91\%  & 89\%\\
					Dec 1 vs Dec 2 & 100\% & 100\% & 100\% & 96\% & 89\% & 84\% & 73\% & 67\% \\
					Dec 1 vs Dec 3 & 100\% & 100\% & 99\% & 96\% & 97\% & 83\% & 72\% & 66\%\\
					Dec 1 vs Dec 4 & 100\% & 100\% & 99\% & 95\% & 85\% & 79\% & 64\% & 57\% \\
					Dec 1 vs Dec 5 & 100\% & 100\% & 99\% & 95\% & 85\% & 79\% & 66\% & 59\% \\
					Dec 1 vs Dec 6 & 100\% & 100\% & 98\% & 94\% & 83\% & 76\% & 62\% & 55\% \\
					Dec 1 vs Dec 7 & 100\% & 100\% & 99\% & 95\% & 82\% & 75\% & 59\% & 51\% \\
					Dec 1 vs Dec 8 & 100\% & 100\% & 99\% & 94\% & 83\% & 78\% & 67\% & 60\% \\
					Dec 1 vs Dec 9 & 100\% & 100\% & 98\% & 94\% & 86\% & 81\% & 73\% & 69\% \\
					Dec 1 vs Dec 10 & 100\% & 100\% & 98\% & 94\% & 88\% & 86\% & 83\% & 82\% \\
					\hline
				\end{tabular}
		\end{threeparttable}}
	\end{table}
	
	\begin{table}[H]
		\centering
		\scalebox{0.9}{
			\begin{threeparttable}
				\caption{Percentage of MVC$\Rightarrow$MEUC for $-(1+Z)^{-a}$}
				\begin{tabular}{ c c c c c c c c c c c}
					\toprule
					& $a=0.01$ & $a=0.3$ & $a=0.5$ & $a=1$ & $a=3$ & $a=5$ & $a=8$ & $a=10$ & $a=15$ & $a=20$\\
					\hline
					Dec 1 vs Dec 1 & 100\% & 100\% & 100\% & 100\% & 98\% & 96\% & 93\% & 91\% & 89\% & 88\% \\
					Dec 1 vs Dec 2 & 100\% & 100\% & 100\% & 99\% & 94\% & 88\% & 78\% & 74\% & 66\% & 63\% \\
					Dec 1 vs Dec 3 & 100\% & 100\% & 100\% & 99\% & 94\% & 87\% & 79\% & 73\% & 66\% & 63\% \\
					Dec 1 vs Dec 4 & 100\% & 100\% & 99\% & 98\% & 93\% & 84\% & 73\% & 65\% & 56\% & 52\% \\
					Dec 1 vs Dec 5 & 100\% & 100\% & 99\% & 99\% & 92\% & 84\% & 74\% & 68\% & 58\% & 54\% \\
					Dec 1 vs Dec 6 & 100\% & 100\% & 99\% & 98\% & 91\% & 82\% & 70\% & 63\% & 55\% & 52\% \\
					Dec 1 vs Dec 7 & 100\% & 100\% & 99\% & 99\% & 91\% & 82\% & 69\% & 62\% & 50\% & 46\% \\
					Dec 1 vs Dec 8 & 99\% & 99\% & 99\% & 98\% & 90\% & 82\% & 72\% & 67\% & 59\% & 56\% \\
					Dec 1 vs Dec 9 & 100\% & 100\% & 99\% & 98\% & 90\% & 84\% & 76\% & 73\% & 68\% & 67\% \\
					Dec 1 vs Dec 10 & 100\% & 99\% & 99\% & 98\% & 91\% & 86\% & 84\% & 83\% & 81\% & 81\% \\
					\hline
				\end{tabular}
		\end{threeparttable}}
	\end{table}
	The results are indicative of the effect of skewness on the decision making of the more risk-averse investors. More specifically, we see that for the logarithmic utility function as well as for $(1+Z)^a$ the MVC looks sufficient for MEUC. Or equivalently, there is evidence that for this specific data both utility functions are almost quadratic. In terms of $-e^{-a(1+Z)}$ and $-(1+Z)^{-a}$ we see that for the more risk-averse cases with $a\geq5$ the skewness plays an important role in the decision making of the investors. Seemingly, one would assume that the results point to the premise of Levy and Markowitz. However, there are two very important factors that we need to highlight. Since the above results originate from real data we have no control on the differences between the means, $\mu_1$ and $\mu_2$, and the differences between the standard deviations, $\sigma_1$ and $\sigma_2$. For example, if we have a closer look on the NYSE statistics per decile, we see that from Decile 8 and on the average standard deviations become markedly larger, which may affect our conclusions. In fact, the larger the differences between the $\mu$ and $\sigma$ parameters, the less the effect of the increased skewness in lottery $Z_2$ on the decision making of investors. The other factor is just a continuation of the previous one. We should not base our conclusions solely on empirical data since by doing that we are unable to cover the entire spectrum of MV-pairs. Specifically, we need an alternative way to deduce in what degree the premise of Levy and Markowitz is satisfied. But still, this simple empirical analysis showcased what we were expecting to see. That some types of investors isnide $\mathbf{U_3}$ will base their decisions more heavily on the relation between the skewnesses regardless if the MVC applies.
	
	Now, as we said, we need to go a step further. Specifically, we are going to use Monte Carlo Simulations in order to have the absolute control in terms of the DGP that generates our data. The simulations not only enable us to choose the type of distribution that generates our data but also enable us to control the levels of differences between the parameters $\mu_1$, $\mu_2$ and $\sigma_1$, $\sigma_2$. Evidently, the larger the differences, $\mu_1/\mu_2$ and $\sigma_1/\sigma_2$ are, the  less the effect of a more skewed or even a more non-normal distribution will be on the MVC's efficiency. Thus, the Monte Carlo Simulations will act as a stress test on the premise of Levy and Markowitz, as we will consider specific cases under which the premise might fail even for less risk-averse investors. In the following section, we aim to measure exactly the efficiency rate of the MVC under some specific cases that we consider.
	
	\subsection{Monte Carlo Simulations}
	The methodology we will follow for the Monte Carlo simulations is analyzed in the following steps. First, we choose the distribution from which we will generate data for two lotteries, $Z_1$ and $Z_2$. 
	\begin{equation*}
		Z_1\sim D(p_1) \quad Z_2\sim D(p_2),
	\end{equation*}
	where $p_1$ and $p_2$ represent the parameters of each distribution.
	
	In our analysis, we use five types of distribution, the Gaussian, the Laplace, the Skew-Normal, the Extreme Value and the Stable Pareto. These are considered good candidates as they are regularly used to fit multiple frequencies of stock returns. The Gaussian as well as Laplace distributions are expected to derive a $100\%$ success of the MVC inferring the MEUC, based on Theorem 2.25. The Skew-Normal represents an interesting case as it belongs to the skew-elliptical family that Schuhmacher et al. \cite{Schuhmacher2021} were referring to. The other two distributions are gradually more skewed and in general more non-normal. So, the last three distributions are considered more interesting. The next step, is to control the differences between the means, the variances and the skewnesses of the two lotteries $Z_1$, $Z_2$. The differences are somewhat based on the differences we found between the parameters in the deciles (see Table 5). Although, we are not bounded to create instances that match real data, since our analysis serves a more general scope. More specifically, we will generate data in such a way that we have absolute control on these differences. This is important as it makes it easier to see the effect of skewness on the investors' decision making. Moreover, the data we generate will always make sure that $Z_1MVZ_2$ and not the other way. We replicate this step multiple times. The data we generate each time are approximately $100,000$ observations. For each distribution and each case of differences in the parameters we generate approximately $1,000$ MV-pairs. So, we can be certain that the findings are robust.
	
	The results produced by the Monte Carlo simulations can be found in the following tables.
	\begin{table}[H]
		\centering
		\scalebox{0.9}{
			\begin{threeparttable}
				\caption{Percentage of MVC$\Rightarrow$MEUC for $(1+Z)^a$}
				\begin{tabular}{c c c c c c}
					\toprule
					Distribution & Parameters Differences & $a=0.01$ & $a=0.1$ & $a=0.5$ & $a=0.9$\\
					\hline
					\multirow{2}{*}{Normal} & $\frac{\mu_1}{\mu_2}=1.05, \ \frac{\sigma_2}{\sigma_1}=1.05$ & 100\% & 100\% & 100\% & 100\% \\
					& $\frac{\mu_1}{\mu_2}=1.01, \ \frac{\sigma_2}{\sigma_1}=1.01$ & 100\% & 100\% & 100\% & 100\% \\
					\hline
					\multirow{2}{*}{Laplace} & $\frac{\mu_1}{\mu_2}=1.05, \ \frac{\sigma_2}{\sigma_1}=1.05$ & 100\% & 100\% & 100\% & 100\% \\
					& $\frac{\mu_1}{\mu_2}=1.01, \ \frac{\sigma_2}{\sigma_1}=1.01$ & 100\% & 100\% & 100\% & 100\% \\
					\hline
					\multirow{4}{*}{SkewN} & $\frac{\mu_1}{\mu_2}=1.05, \ \frac{\sigma_2}{\sigma_1}=1.05, \ \frac{s_2}{s_1}=1.5$ & 100\% & 100\% & 100\% & 100\% \\
					& $\frac{\mu_1}{\mu_2}=1.05, \ \frac{\sigma_2}{\sigma_1}=1.05, \ \frac{s_2}{s_1}=3$ & 100\% & 100\% & 100\% & 100\% \\
					& $\frac{\mu_1}{\mu_2}=1.01, \ \frac{\sigma_2}{\sigma_1}=1.01, \ \frac{s_2}{s_1}=1.5$ & 100\% & 100\% & 100\% & 100\% \\
					& $\frac{\mu_1}{\mu_2}=1.01, \ \frac{\sigma_2}{\sigma_1}=1.01, \ \frac{s_2}{s_1}=3$ & 100\% & 100\% & 100\%& 100\%  \\
					\hline
					\multirow{4}{*}{Extreme} & $\frac{\mu_1}{\mu_2}=1.05, \ \frac{\sigma_2}{\sigma_1}=1.05, \ \frac{s_2}{s_1}=1.5$ & 100\% & 100\% & 100\% & 100\% \\
					& $\frac{\mu_1}{\mu_2}=1.05, \ \frac{\sigma_2}{\sigma_1}=1.05, \ \frac{s_2}{s_1}=3$ & 100\% & 100\% & 100\% & 100\% \\
					& $\frac{\mu_1}{\mu_2}=1.01, \ \frac{\sigma_2}{\sigma_1}=1.01, \ \frac{s_2}{s_1}=1.5$ & 100\% & 100\% & 100\% & 100\% \\
					& $\frac{\mu_1}{\mu_2}=1.01, \ \frac{\sigma_2}{\sigma_1}=1.01, \ \frac{s_2}{s_1}=3$ & 81\% & 86\% & 91\% & 94\% \\
					\hline
					\multirow{3}{*}{Stable} & $1.3<\frac{\mu_1}{\mu_2}\leq1.5, \ 1.3<\frac{\sigma_2}{\sigma_1}\leq1.5, \ 1.5\leq\frac{s_2}{s_1}\leq3$ & 100\% & 100\% & 100\% & 100\% \\
					& $1.1<\frac{\mu_1}{\mu_2}\leq1.3, \ 1.1<\frac{\sigma_2}{\sigma_1}\leq1.3, \ 1.5\leq\frac{s_2}{s_1}\leq3$ & 97\% & 98\% & 100\% & 100\% \\
					& $1.01\leq\frac{\mu_1}{\mu_2}\leq1.1, \ 1.01\leq\frac{\sigma_2}{\sigma_1}\leq1.1, \ 1.5\leq\frac{s_2}{s_1}\leq3$ &  67\% & 69\% & 85\% & 97\% \\
					\hline
				\end{tabular}
		\end{threeparttable}}
	\end{table}

	\begin{table}[H]
		\centering
		\scalebox{0.9}{
			\begin{threeparttable}
				\caption{Percentage of MVC$\Rightarrow$MEUC for $log(a+Z)$}
				\begin{tabular}{c c c c}
					\toprule
					Distribution & Parameters Differences & $a=0.9$ & $a=1$ \\
					\hline
					\multirow{2}{*}{Normal} & $\frac{\mu_1}{\mu_2}=1.05, \ \frac{\sigma_2}{\sigma_1}=1.05$ & 100\% & 100\% \\
					& $\frac{\mu_1}{\mu_2}=1.01, \ \frac{\sigma_2}{\sigma_1}=1.01$ & 100\% & 100\% \\
					\hline
					\multirow{2}{*}{Laplace} & $\frac{\mu_1}{\mu_2}=1.05, \ \frac{\sigma_2}{\sigma_1}=1.05$ & 100\% & 100\% \\
					& $\frac{\mu_1}{\mu_2}=1.01, \ \frac{\sigma_2}{\sigma_1}=1.01$ & 100\% & 100\% \\
					\hline
					\multirow{4}{*}{SkewN} & $\frac{\mu_1}{\mu_2}=1.05, \ \frac{\sigma_2}{\sigma_1}=1.05, \ \frac{s_2}{s_1}=1.5$ & 100\% & 100\% \\
					& $\frac{\mu_1}{\mu_2}=1.05, \ \frac{\sigma_2}{\sigma_1}=1.05, \ \frac{s_2}{s_1}=3$ & 100\% & 100\% \\
					& $\frac{\mu_1}{\mu_2}=1.01, \ \frac{\sigma_2}{\sigma_1}=1.01, \ \frac{s_2}{s_1}=1.5$ & 100\% & 100\% \\
					& $\frac{\mu_1}{\mu_2}=1.01, \ \frac{\sigma_2}{\sigma_1}=1.01, \ \frac{s_2}{s_1}=3$ & 100\% & 100\% \\
					\hline
					\multirow{4}{*}{Extreme} & $\frac{\mu_1}{\mu_2}=1.05, \ \frac{\sigma_2}{\sigma_1}=1.05, \ \frac{s_2}{s_1}=1.5$ & 100\% & 100\% \\
					& $\frac{\mu_1}{\mu_2}=1.05, \ \frac{\sigma_2}{\sigma_1}=1.05, \ \frac{s_2}{s_1}=3$ & 100\% & 100\% \\
					& $\frac{\mu_1}{\mu_2}=1.01, \ \frac{\sigma_2}{\sigma_1}=1.01, \ \frac{s_2}{s_1}=1.5$ & 100\% & 100\% \\
					& $\frac{\mu_1}{\mu_2}=1.01, \ \frac{\sigma_2}{\sigma_1}=1.01, \ \frac{s_2}{s_1}=3$ & 81\% & 85\% \\
					\hline
					\multirow{3}{*}{Stable} & $1.3<\frac{\mu_1}{\mu_2}\leq1.5, \ 1.3<\frac{\sigma_2}{\sigma_1}\leq1.5, \ 1.5\leq\frac{s_2}{s_1}\leq3$ & 100\% & 100\% \\
					& $1.1<\frac{\mu_1}{\mu_2}\leq1.3, \ 1.1<\frac{\sigma_2}{\sigma_1}\leq1.3, \ 1.5\leq\frac{s_2}{s_1}\leq3$ & 92\% & 96\% \\
					& $1.01\leq\frac{\mu_1}{\mu_2}\leq1.1, \ 1.01\leq\frac{\sigma_2}{\sigma_1}\leq1.1, \ 1.5\leq\frac{s_2}{s_1}\leq3$ &  63\% & 67\% \\
					\hline
				\end{tabular}
		\end{threeparttable}}
	\end{table}

	\begin{table}[H]
		\centering
		\scalebox{0.6}{
			\begin{threeparttable}
				\caption{Percentage of MVC$\Rightarrow$MEUC for $-e^{-a(1+Z)}$}
				\begin{tabular}{c c c c c c c c c c}
					\toprule
					Distribution & Parameters Differences & $a=0.7$ & $a=1$ & $a=3$ & $a=5$ & $a=8$ & $a=10$ & $a=15$ & $a=20$ \\
					\hline
					\multirow{2}{*}{Normal} & $\frac{\mu_1}{\mu_2}=1.05, \ \frac{\sigma_2}{\sigma_1}=1.05$ & 100\% & 100\% & 100\% & 100\% & 100\% & 100\% & 100\% & 100\% \\
					& $\frac{\mu_1}{\mu_2}=1.01, \ \frac{\sigma_2}{\sigma_1}=1.01$ & 100\% & 100\% & 100\% & 100\% & 100\% & 100\% & 100\% & 100\% \\
					\hline
					\multirow{2}{*}{Laplace} & $\frac{\mu_1}{\mu_2}=1.05, \ \frac{\sigma_2}{\sigma_1}=1.05$ & 100\% & 100\% & 100\% & 100\% & 100\% & 100\% & 100\% & 100\% \\
					& $\frac{\mu_1}{\mu_2}=1.01, \ \frac{\sigma_2}{\sigma_1}=1.01$ & 100\% & 100\% & 100\% & 100\% & 100\% & 100\% & 100\% & 100\% \\
					\hline
					\multirow{4}{*}{SkewN} & $\frac{\mu_1}{\mu_2}=1.05, \ \frac{\sigma_2}{\sigma_1}=1.05, \ \frac{s_2}{s_1}=1.5$ & 100\% & 100\% & 100\% & 100\% & 100\% & 100\% & 100\% & 100\% \\
					& $\frac{\mu_1}{\mu_2}=1.05, \ \frac{\sigma_2}{\sigma_1}=1.05, \ \frac{s_2}{s_1}=3$ & 100\% & 100\% & 100\% & 100\% & 100\% & 100\% & 100\% & 100\% \\
					& $\frac{\mu_1}{\mu_2}=1.01, \ \frac{\sigma_2}{\sigma_1}=1.01, \ \frac{s_2}{s_1}=1.5$ & 100\% & 100\% & 100\% & 100\% & 100\% & 100\% & 91\% & 18\% \\
					& $\frac{\mu_1}{\mu_2}=1.01, \ \frac{\sigma_2}{\sigma_1}=1.01, \ \frac{s_2}{s_1}=3$ & 100\% & 100\%  & 100\% & 100\% & 100\% & 100\% & 4\% & 0\% \\
					\hline
					\multirow{4}{*}{Extreme} & $\frac{\mu_1}{\mu_2}=1.05, \ \frac{\sigma_2}{\sigma_1}=1.05, \ \frac{s_2}{s_1}=1.5$ & 100\% & 100\% & 100\% & 100\% & 100\% & 100\% & 100\% & 100\% \\
					& $\frac{\mu_1}{\mu_2}=1.05, \ \frac{\sigma_2}{\sigma_1}=1.05, \ \frac{s_2}{s_1}=3$ & 100\% & 100\% & 100\% & 100\% & 6\% & 0\% & 0\% & 0\% \\
					& $\frac{\mu_1}{\mu_2}=1.01, \ \frac{\sigma_2}{\sigma_1}=1.01, \ \frac{s_2}{s_1}=1.5$ & 100\% & 100\% & 99\% & 80\% & 7\% & 0\% & 0\% & 0\% \\
					& $\frac{\mu_1}{\mu_2}=1.01, \ \frac{\sigma_2}{\sigma_1}=1.01, \ \frac{s_2}{s_1}=3$ & 94\% & 49\% & 1\% & 0\% & 0\% & 0\% & 0\% & 0\% \\
					\hline
					\multirow{3}{*}{Stable} & $1.3<\frac{\mu_1}{\mu_2}\leq1.5, \ 1.3<\frac{\sigma_2}{\sigma_1}\leq1.5, \ 1.5\leq\frac{s_2}{s_1}\leq3$ & 100\% & 100\% & 99\% & 94\% & 88\% & 83\% & 78\% & 74\% \\
					& $1.1<\frac{\mu_1}{\mu_2}\leq1.3, \ 1.1<\frac{\sigma_2}{\sigma_1}\leq1.3, \ 1.5\leq\frac{s_2}{s_1}\leq3$ & 100\% & 100\% & 78\% & 62\% & 51\% & 45\% & 43\% & 42\% \\
					& $1.01\leq\frac{\mu_1}{\mu_2}\leq1.1, \ 1.01\leq\frac{\sigma_2}{\sigma_1}\leq1.1, \ 1.5\leq\frac{s_2}{s_1}\leq3$ & 95\% & 85\% & 44\% & 26\% & 28\% & 28\% & 28\% & 26\% \\
					\hline
				\end{tabular}
		\end{threeparttable}}
	\end{table}
	
	\begin{table}[H]
		\centering
		\scalebox{0.6}{
			\begin{threeparttable}
				\caption{Percentage of MVC$\Rightarrow$MEUC for $-(1+Z)^{-a}$}
				\begin{tabular}{c c c c c c c c c c c c}
					\toprule
					Distribution & Parameters Differences & $a=0.01$ & $a=0.3$ & $a=0.5$ & $a=1$ & $a=3$ & $a=5$ & $a=8$ & $a=10$ & $a=15$ & $a=20$ \\
					\hline
					\multirow{2}{*}{Normal} & $\frac{\mu_1}{\mu_2}=1.05, \ \frac{\sigma_2}{\sigma_1}=1.05$ & 100\% & 100\% & 100\% & 100\% & 100\% & 100\% & 100\% & 100\% & 100\% & 100\% \\
					& $\frac{\mu_1}{\mu_2}=1.01, \ \frac{\sigma_2}{\sigma_1}=1.01$ & 100\% & 100\% & 100\% & 100\% & 100\% & 100\% & 100\% & 100\% & 100\% & 100\% \\
					\hline
					\multirow{2}{*}{Laplace} & $\frac{\mu_1}{\mu_2}=1.05, \ \frac{\sigma_2}{\sigma_1}=1.05$ & 100\% & 100\% & 100\% & 100\% & 100\% & 100\% & 100\% & 100\% & 100\% & 100\% \\
					& $\frac{\mu_1}{\mu_2}=1.01, \ \frac{\sigma_2}{\sigma_1}=1.01$ & 100\% & 100\% & 100\% & 100\% & 100\% & 100\% & 100\% & 100\% & 100\% & 100\% \\
					\hline
					\multirow{4}{*}{SkewN} & $\frac{\mu_1}{\mu_2}=1.05, \ \frac{\sigma_2}{\sigma_1}=1.05, \ \frac{s_2}{s_1}=1.5$ & 100\% & 100\% & 100\% & 100\% & 100\% & 100\% & 100\% & 100\% & 100\% & 100\% \\
					& $\frac{\mu_1}{\mu_2}=1.05, \ \frac{\sigma_2}{\sigma_1}=1.05, \ \frac{s_2}{s_1}=3$ & 100\% & 100\% & 100\% & 100\% & 100\% & 100\% & 100\% & 100\% & 100\% & 100\% \\
					& $\frac{\mu_1}{\mu_2}=1.01, \ \frac{\sigma_2}{\sigma_1}=1.01, \ \frac{s_2}{s_1}=1.5$ & 100\% & 100\% & 100\% & 100\% & 100\% & 100\% & 100\% & 100\% & 82\% & 15\% \\
					& $\frac{\mu_1}{\mu_2}=1.01, \ \frac{\sigma_2}{\sigma_1}=1.01, \ \frac{s_2}{s_1}=3$ & 100\% & 100\% & 100\%& 100\%  & 100\% & 100\% & 100\% & 76\% & 0\% & 0\% \\
					\hline
					\multirow{4}{*}{Extreme} & $\frac{\mu_1}{\mu_2}=1.05, \ \frac{\sigma_2}{\sigma_1}=1.05, \ \frac{s_2}{s_1}=1.5$ & 100\% & 100\% & 100\% & 100\% & 100\% & 100\% & 100\% & 100\% & 100\% & 100\% \\
					& $\frac{\mu_1}{\mu_2}=1.05, \ \frac{\sigma_2}{\sigma_1}=1.05, \ \frac{s_2}{s_1}=3$ & 100\% & 100\% & 100\% & 100\% & 100\% & 100\% & 0\% & 0\% & 0\% & 0\% \\
					& $\frac{\mu_1}{\mu_2}=1.01, \ \frac{\sigma_2}{\sigma_1}=1.01, \ \frac{s_2}{s_1}=1.5$ & 100\% & 100\% & 100\% & 100\% & 92\% & 49\% & 0\% & 0\% & 0\% & 0\% \\
					& $\frac{\mu_1}{\mu_2}=1.01, \ \frac{\sigma_2}{\sigma_1}=1.01, \ \frac{s_2}{s_1}=3$ & 72\% & 68\% & 55\% & 37\% & 0\% & 0\% & 0\% & 0\% & 0\% & 0\% \\
					\hline
					\multirow{3}{*}{Stable} & $1.3<\frac{\mu_1}{\mu_2}\leq1.5, \ 1.3<\frac{\sigma_2}{\sigma_1}\leq1.5, \ 1.5\leq\frac{s_2}{s_1}\leq3$ & 99\% & 99\% & 97\% & 97\% & 93\% & 85\% & 81\% & 79\% & 75\% & 73\% \\
					& $1.1<\frac{\mu_1}{\mu_2}\leq1.3, \ 1.1<\frac{\sigma_2}{\sigma_1}\leq1.3, \ 1.5\leq\frac{s_2}{s_1}\leq3$ & 90\% & 88\% & 87\% & 85\% & 61\% & 51\% & 46\% & 43\% & 42\% & 41\% \\
					& $1.01\leq\frac{\mu_1}{\mu_2}\leq1.1, \ 1.01\leq\frac{\sigma_2}{\sigma_1}\leq1.1, \ 1.5\leq\frac{s_2}{s_1}\leq3$ &  62\% & 59\% & 59\% & 51\% & 28\% & 28\% & 28\% & 28\% & 26\% & 23\% \\
					\hline
				\end{tabular}
		\end{threeparttable}}
	\end{table}
	
	Before getting into the results we need to highlight some important features of our generated data. Notice that for the Skew-Normal as well as the Extreme Value distribution the differences between the means and standard deviations we consider are small as we found out that for larger differences the MVC works fine. On the contrary, the Stable Pareto distribution allows the distances between means and standard deviations to go as high as $50\%$. This is due to the extreme characteristics of the Stable Pareto distribution, which is known to exhibit sudden large jumps and thus creates higher challenges for the MVC. In terms of the Stable Pareto distribution, the reason we allow the differences between the parameters to move inside a specific range is because it is harder to control the data produced, since the distribution has undefined moments. 
	
	Now, the Monte Carlo Simulations indicate that the $(1+Z)^a$ and $\log(a+Z)$ utility functions are only mildly affected by the Skew-Normal and Extreme Value distributions when the differences between the parameters are very close in value. This shows, that the less risk averse investors can generally trust the MVC for their decision making, even for mildly non-normal skewed distributions. But, in the Stable Pareto case when the differences between the means and the standard deviations hover around $1\%$ and $10\%$ the investors should take into consideration the skewnesses of the two generated processes in order to make better decisions. With regards to $-e^{-a(1+Z)}$ and $-(1+Z)^{-a}$ the issues with the sufficiency of the MVC for the MEUC are evident even in the case of the Skew-Normal distribution. More specifically, the extremely risk-averse investors with $a=15,20$ will make very wrong decisions when the means and standard deviations differences are very close in value. The results are even worse for the Extreme Value distribution case. In particular, we see that the impact on the MVC is evident even for less risk-averse investors and is far worse as the differences in the parameters get closer. But the more interesting results come from the Stable Pareto case, which signifies that these types of investors will need more information on the lotteries' characteristics, besides the means and variances, in order to make their decision. This is evident even in cases where the differences between the parameters get as high as $30\%$.
	
	To sum up, we conclude that the main issue with the premise of Levy and Markowitz lies with the more non-normal cases. These cases can better be described by the Extreme Value and Stable Pareto distributions. However, under the assumption of Gaussian, Laplace or Skew-Normal distributions the premise of Levy and Markowitz seems valid for all the utility functions that we put to test. In other words, under those distributions the above utility functions are almost quadratic. The Extreme Value distribution represents the first step in testing the premise under more non-normal cases. The MVC seems to resist the pressure for less risk-averse investors. The more noteworthy findings are derived by the Stable Pareto distribution, which, according to Mandelbrot, represents an appropriate description of the movement of daily stock returns. In this case, letting the means' and standard deviations' differences hover around $1\%$ and $30\%$ results in erroneous decisions made by almost all parametrizations of the four investors. So, as expected, the greater the non-normality of the distribution of lotteries, the more information the decision makers will need, with respect to the characteristics of the distribution of each lottery, in order to make the right decision. As a result, we conclude that the premise of Levy and Markowitz works appropriately for Elliptical or Skew-Elliptical distributions. However, as we have thoroughly discussed, under the elliptical family of distributions the MVC is equivalent to the MEUC, for any concave utility function. So, the additional assumption of approximately quadratic utility functions is unnecessary. With regards to the Skew-Elliptical family of distributions we find some extreme cases for which only the very risk-averse investors might need to know the level of skewness of each lottery. Lastly, when departing from normality the premise of Levy and Markowitz is problematic for all the utility functions that we put to test. At this point we may give an answer to our initial question concerning the size of $\mathbf{U^*_3}$. In particular, as we have shown, the Elliptical family of distributions is expected to deliver MVC$\Rightarrow$MEUC for the four utility functions we put to test, without the need of the utility functions being approximately quadratic. So, such cases should not be considered to comprise $\mathbf{U^*_3}$. On the other hand, under the more non-normal distributions the four utility functions cannot be included in $\mathbf{U^*_3}$.
	
	\section{Conclusions}
	Since its conception in 1952, the MVC has gone through an extensive amount of criticism when it comes to its realism and usefulness. The main argument has always been that the underlying assumptions of either (i) quadratic preferences, or (ii) Gaussian distributions, are unrealistic. Markowitz (2010, 2014) \cite{Markowitz2010} and \cite{Markowitz2014} insists that the literature has misinterpreted his model. This stimulated us to revisit the MVC, so as to examine Markowitz's remark as well as to clarify how it is associated with the SD rules.
	
	We analyzed thoroughly the literature to clarify which are the conditions that make the MVC coincide with the MEUC. We found that the elliptical family of distributions can replace the assumption of normality, based on the findings of Chamberlain (1983). However, the more recent findings of Schuhmacher et al. (2021) with respect to the Skew-Elliptical family are not consistent with our empirical results. In particular, we saw that under Skew-Normal distributions there are very-risk averse investors that are part of $\mathbf{U_3}$, that may require to know the level of skewness of each lottery. Thus, we concluded that under Skew-Elliptical distributions we cannot claim that the MVC coincides with the MEUC.
	
	From there, we went on to identify the class of investors for which the MVC coincides with the MEUC. We investigated Markowitz's premise in 1959, arguing that we only need approximately quadratic utility functions to make the MVC equivalent to the MEUC. We argued that the evidence from \cite{Markowitz1959} and \cite{Levy1979} does not suffice to support that premise. In fact, we proposed the use of Monte Carlo simulations in order to test Markowitz's premise under multiple types of distributions, for a specific choice of investors. We found out that under a Skew-Normal distribution, the MVC is equivalent to the MEUC except for the very risk-averse investors. But for more non-normal distributions, like the Extreme Value and Stable Pareto, even less risk-averse investors will come to a high percentage of wrong decisions if they use only the information coming from the MVC. Thus, based on our findings, we deduced that Markowitz's premise seems to work for Skew-Elliptical distributions, but this is not the case for more non-normal distributions.
	
	\newpage
	\section{Appendix}
	\begin{figure}[H]
		\caption{$(1+Z)^a$ vs its Quadratic approximation around $0$, for $Z\in[-0.9,1]$}
		\begin{subfigure}{0.45\textwidth}	
			\includegraphics[width=\linewidth]{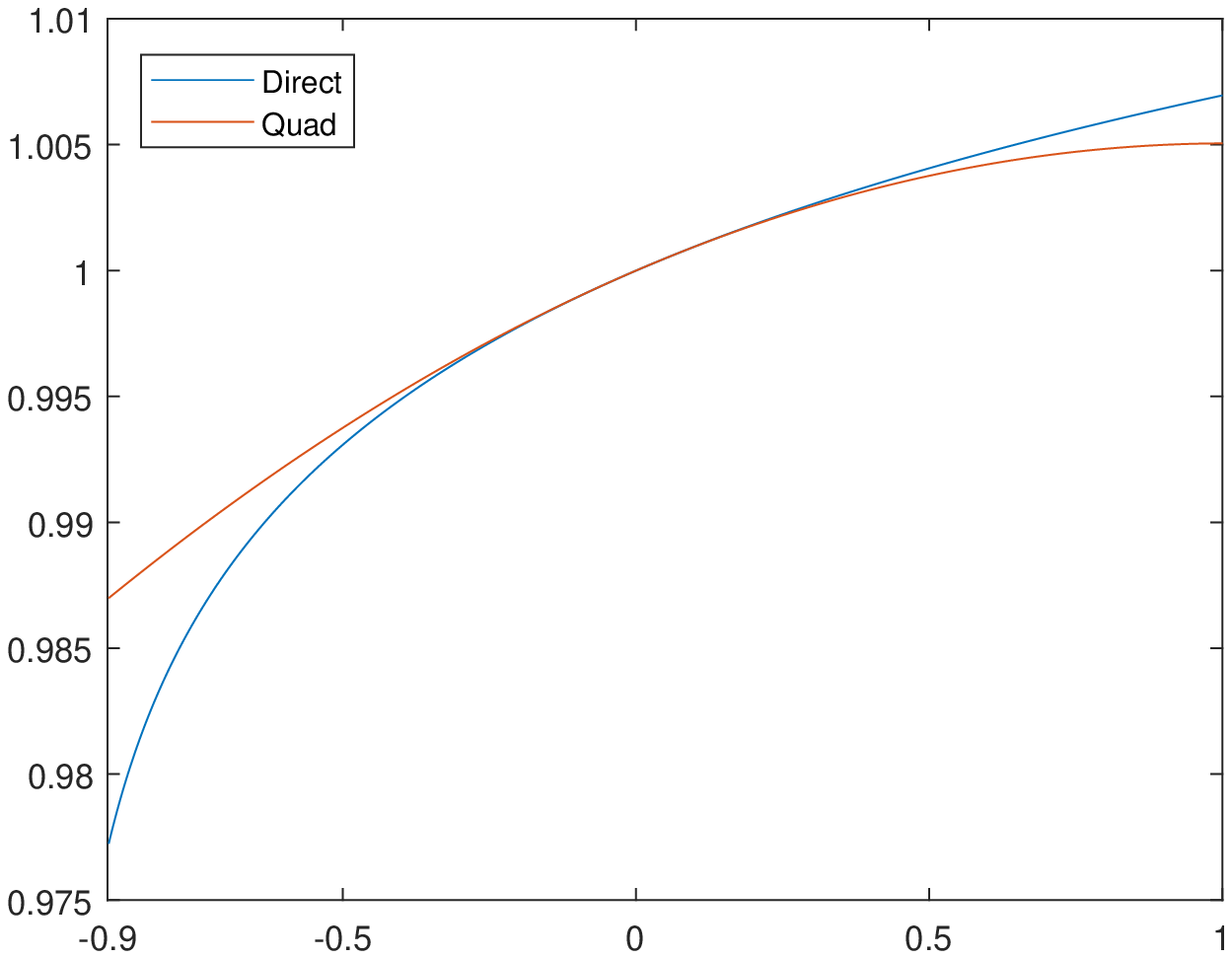}
			\caption{$a=0.01$}
		\end{subfigure}
		\hspace*{1in}
		\begin{subfigure}{0.45\textwidth}
			\includegraphics[width=\linewidth]{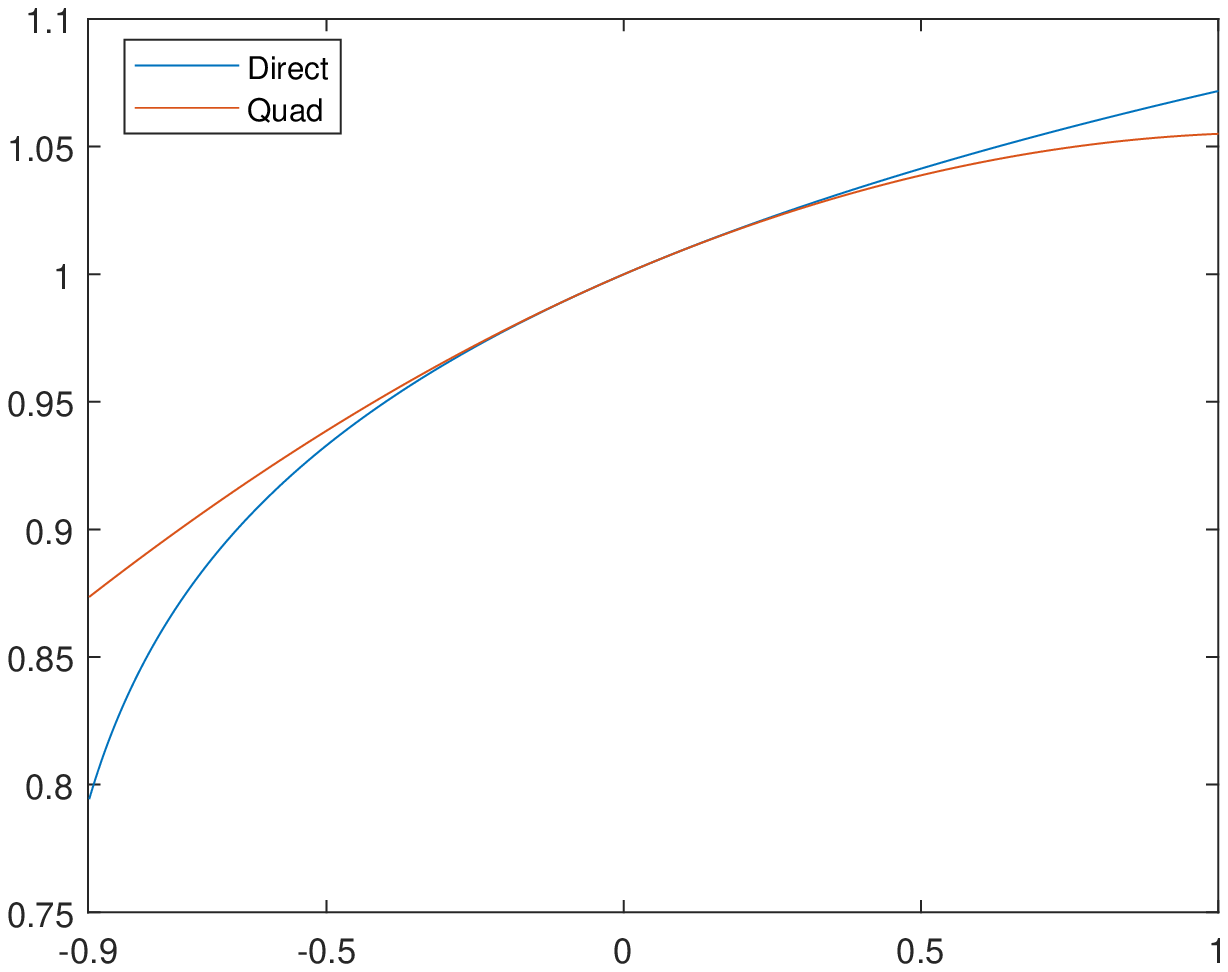}
			\caption{$a=0.10$}
		\end{subfigure}
		\begin{subfigure}{0.45\textwidth}
			\includegraphics[width=\linewidth]{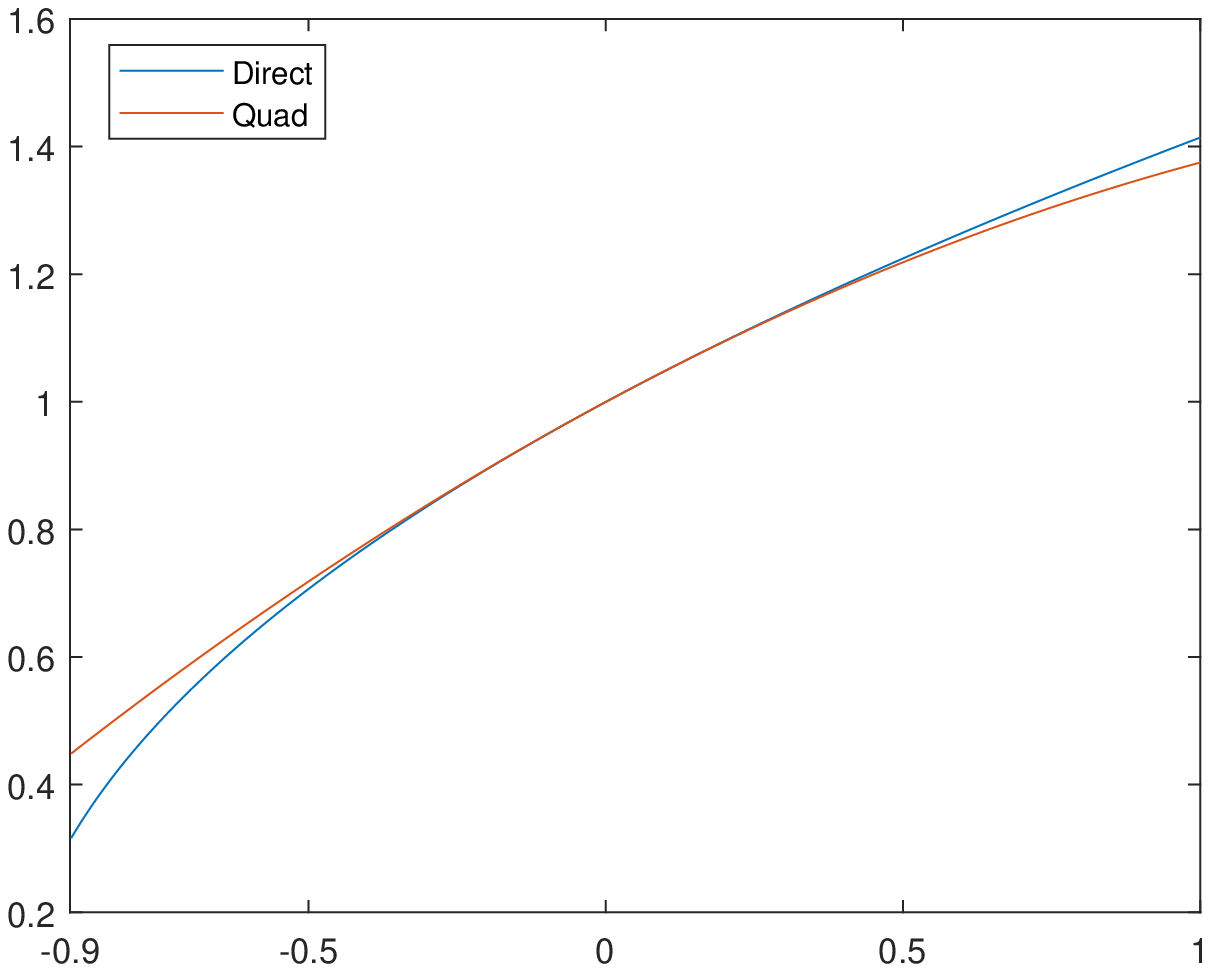}
			\caption{$a=0.50$}
		\end{subfigure}
		\hspace*{1in}
		\begin{subfigure}{0.45\textwidth}
			\includegraphics[width=\linewidth]{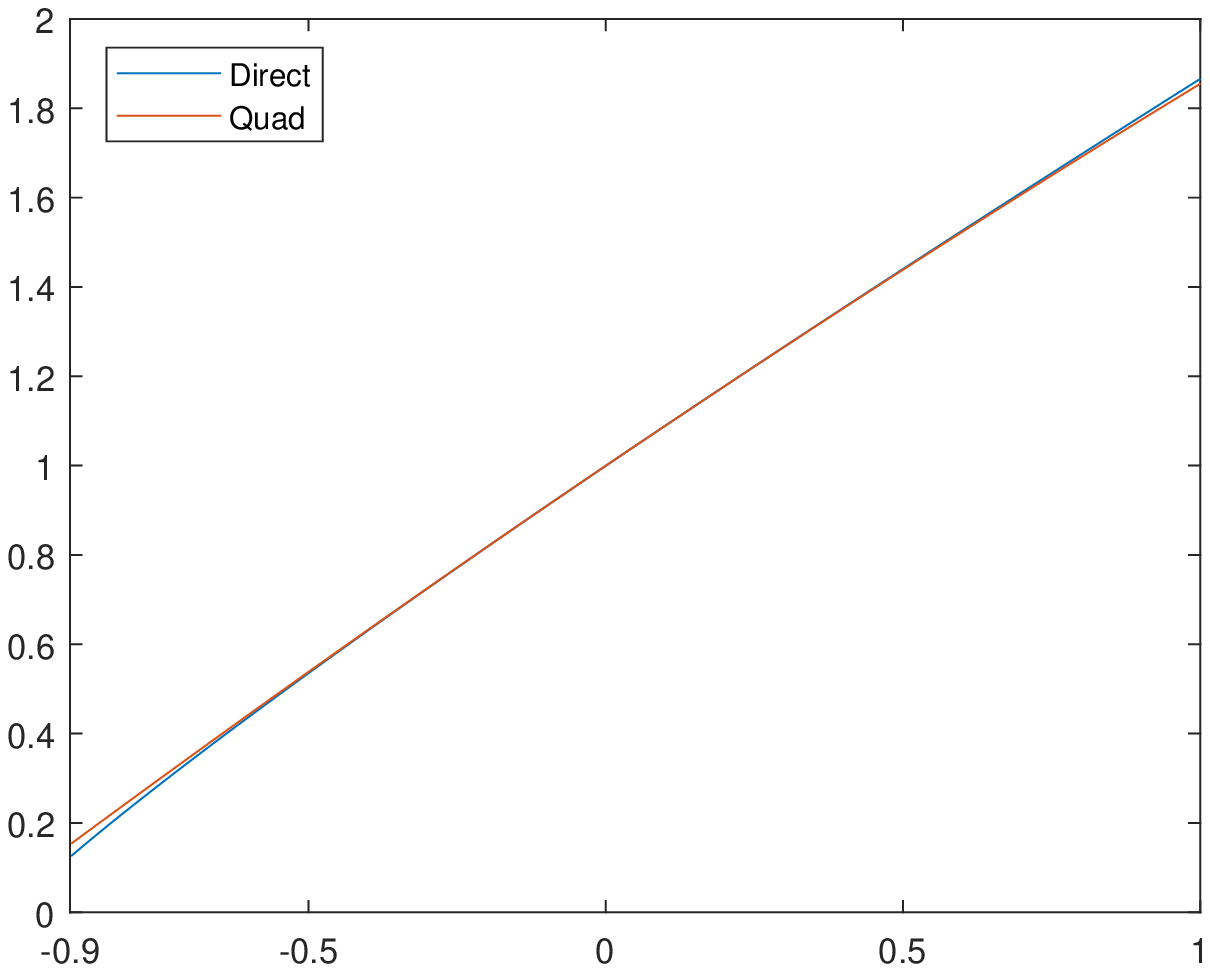}
			\caption{$a=0.90$}
		\end{subfigure}
	\end{figure}
	
	\newpage
	\begin{figure}[H]
		\caption{$log(a+Z)$ vs its Quadratic approximation around $0$, for $Z\in[-0.9,1]$}
		\begin{subfigure}{0.45\textwidth}
			\includegraphics[width=\linewidth]{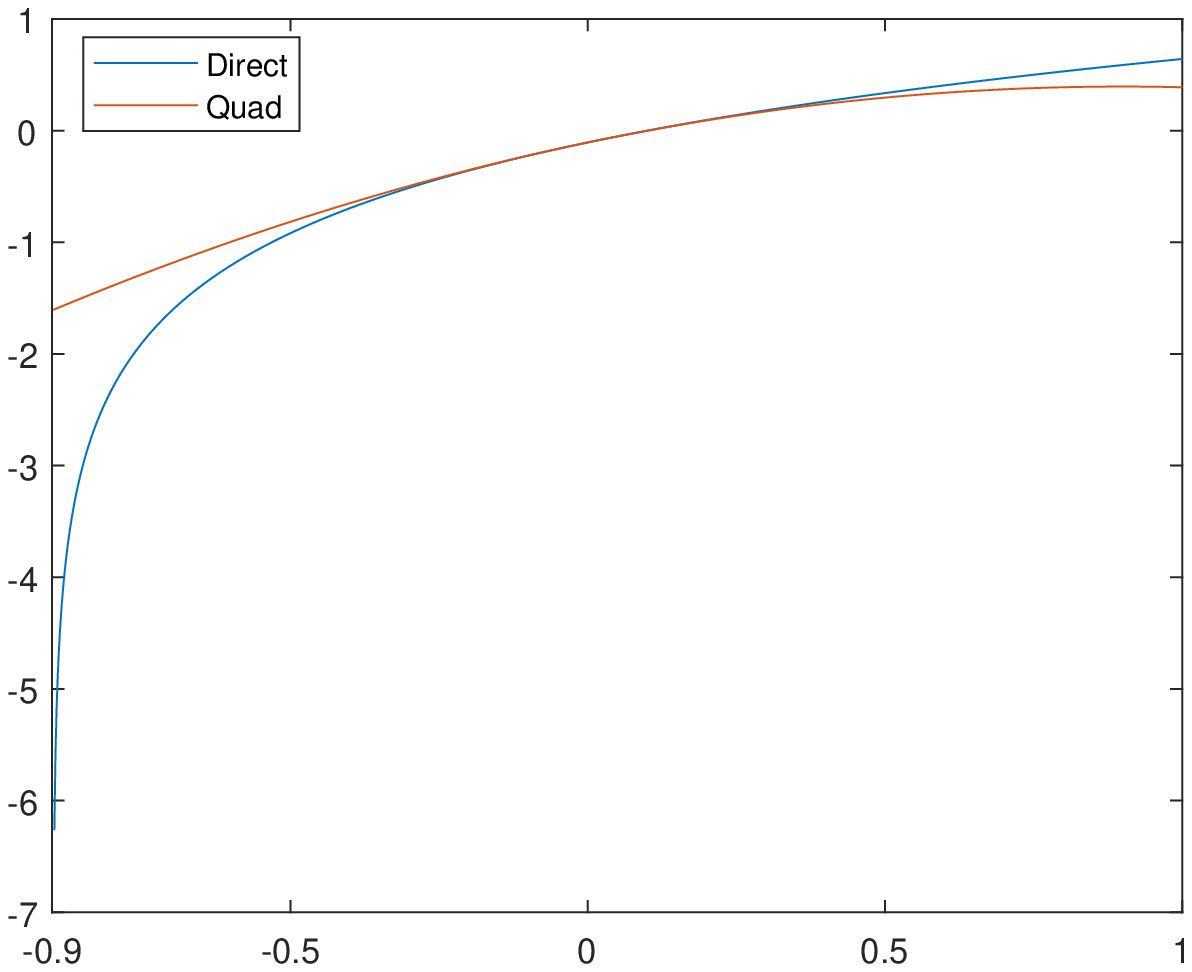}
			\caption{$a=0.90$}
		\end{subfigure}
		\hspace*{1in}
		\begin{subfigure}{0.45\textwidth}
			\includegraphics[width=\linewidth]{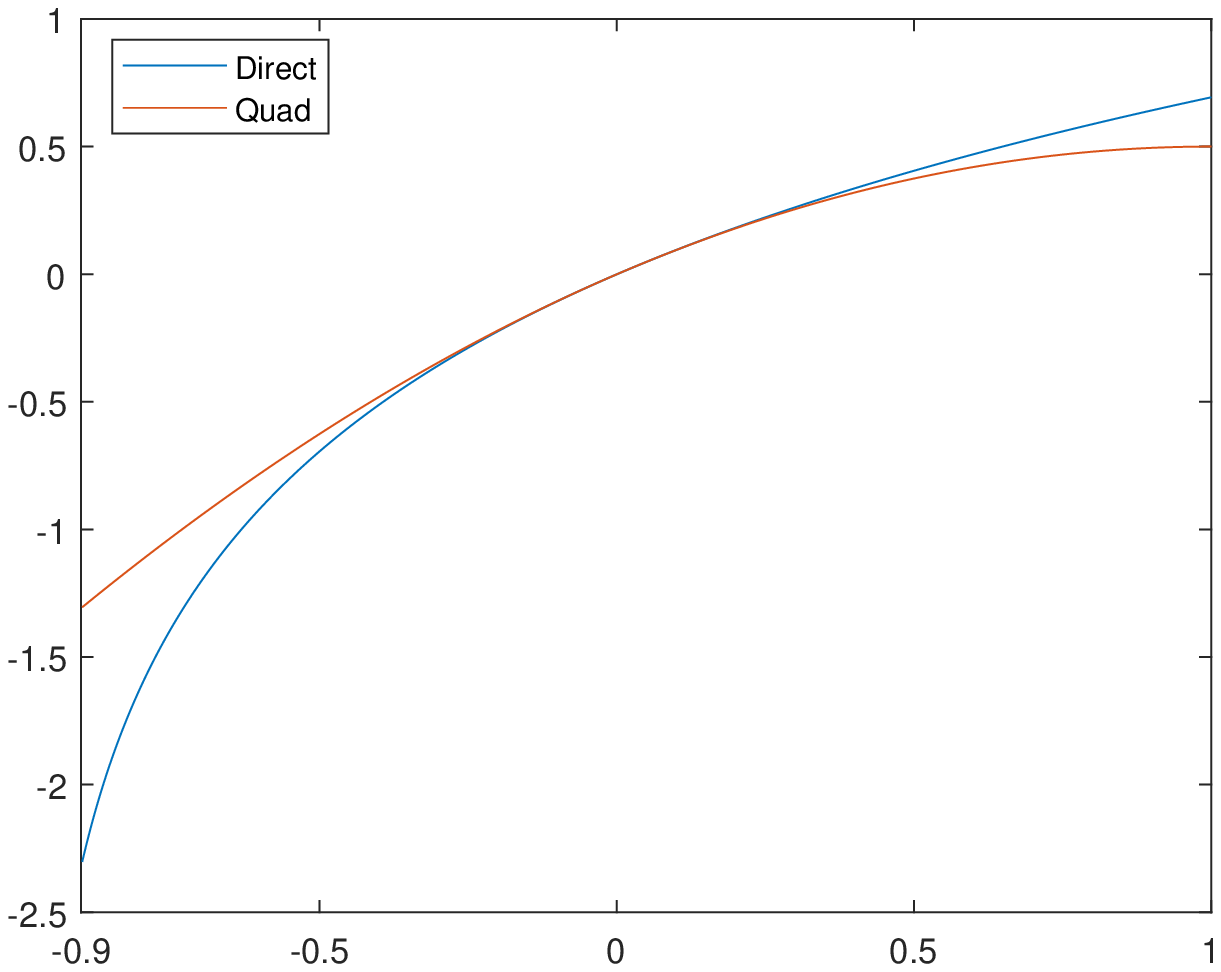}
			\caption{$a=1$}
		\end{subfigure}
	\end{figure}
	
	\newpage
	\begin{figure}[H]
		\caption{$-e^{-a(1+Z)}$ vs its Quadratic approximation around $0$, for $Z\in[-0.9,1]$}
		\begin{subfigure}{0.45\textwidth}
			\includegraphics[width=\linewidth]{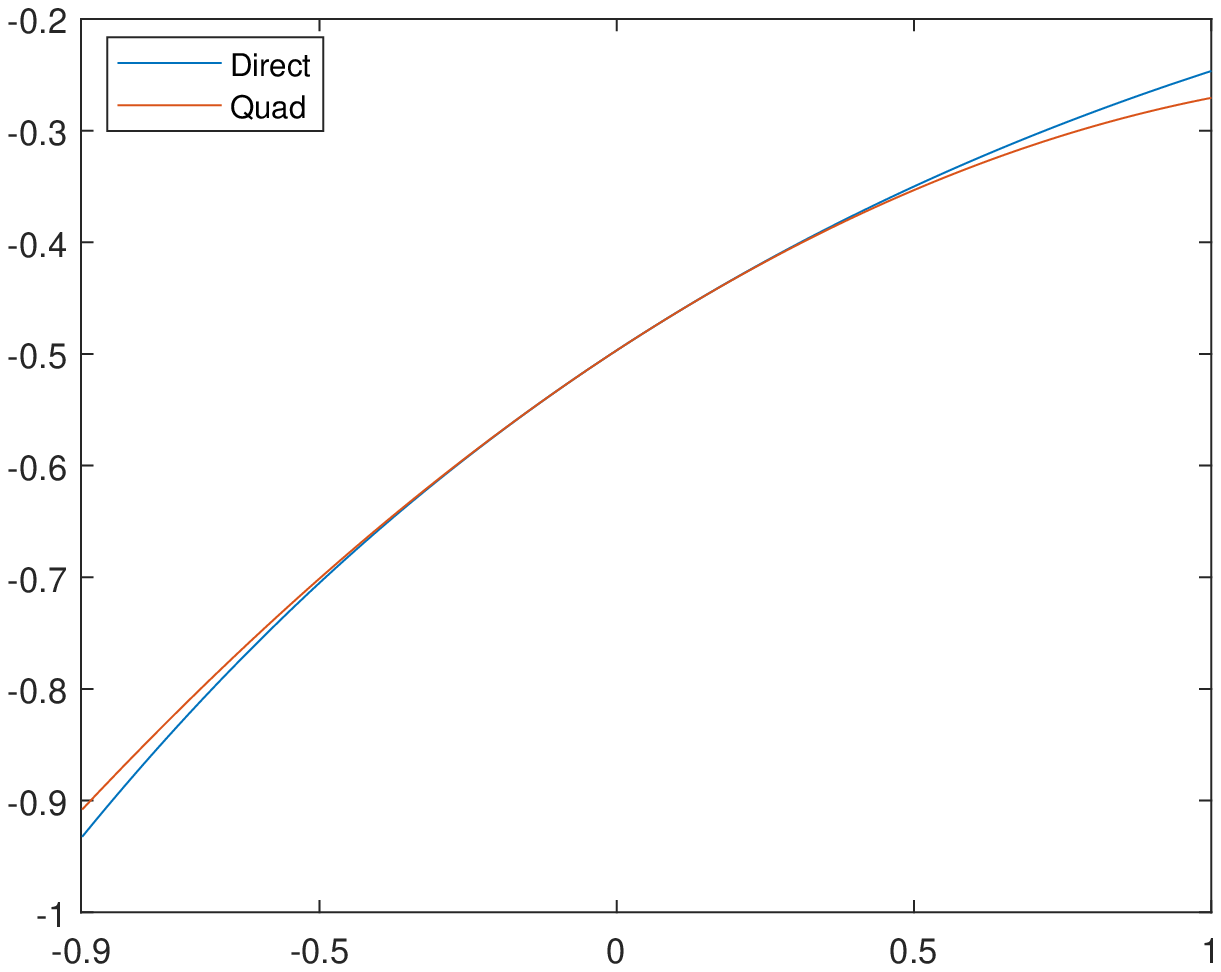}
			\caption{$a=0.70$}
		\end{subfigure}
		\hspace*{1in}
		\begin{subfigure}{0.45\textwidth}
			\includegraphics[width=\linewidth]{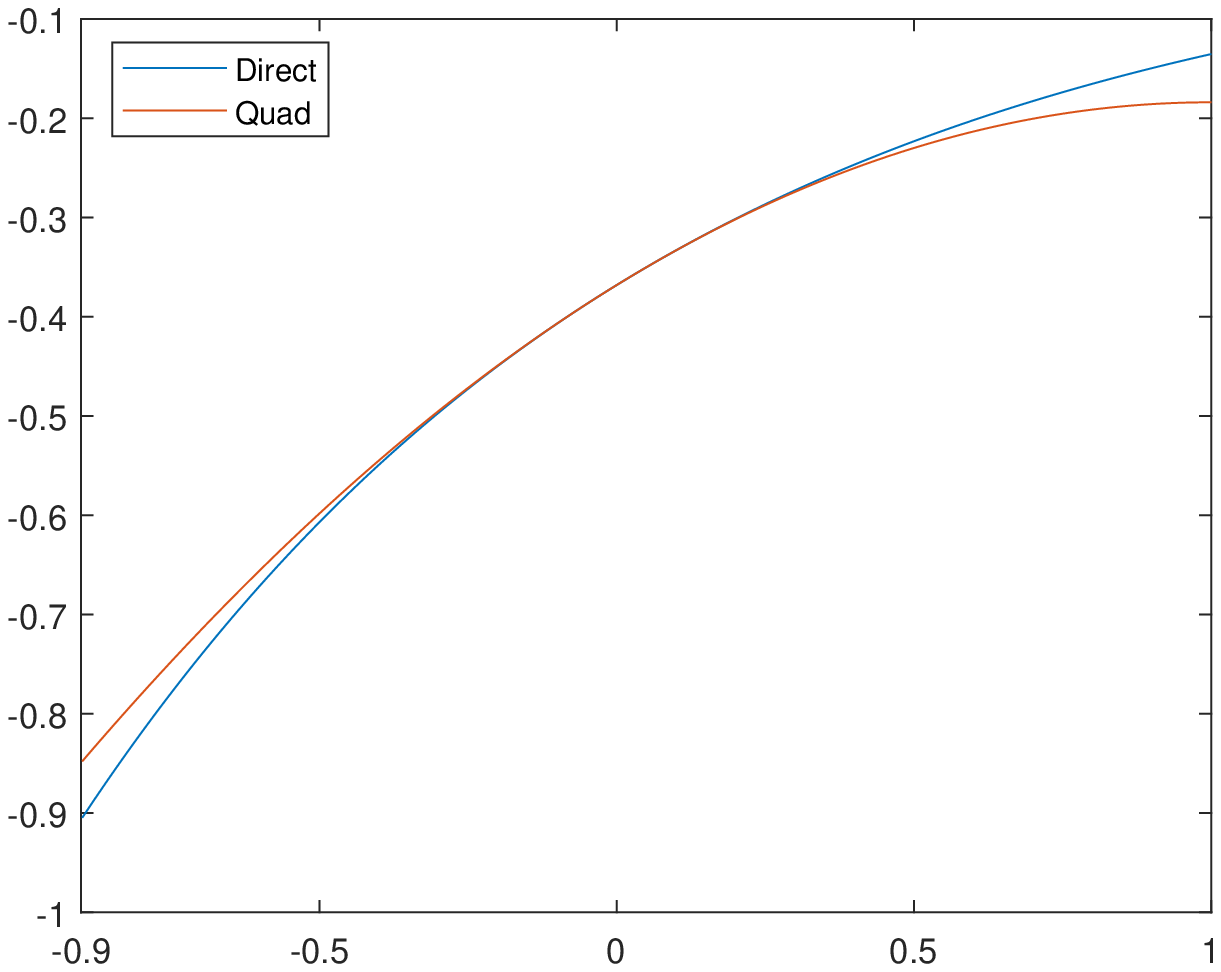}
			\caption{$a=1$}
		\end{subfigure}
		\begin{subfigure}{0.45\textwidth}
			\includegraphics[width=\linewidth]{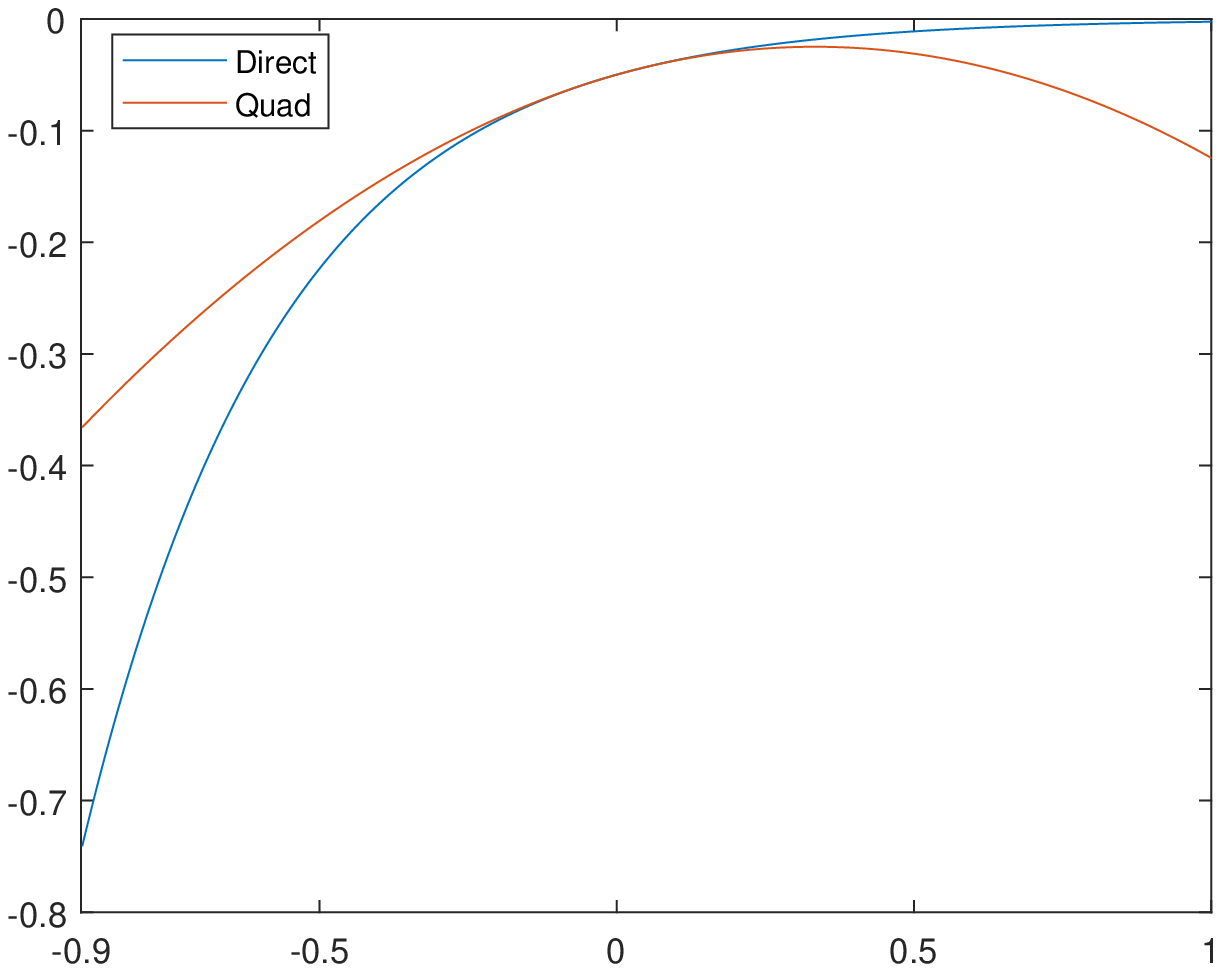}
			\caption{$a=3$}
		\end{subfigure}
		\hspace*{1in}
		\begin{subfigure}{0.45\textwidth}
			\includegraphics[width=\linewidth]{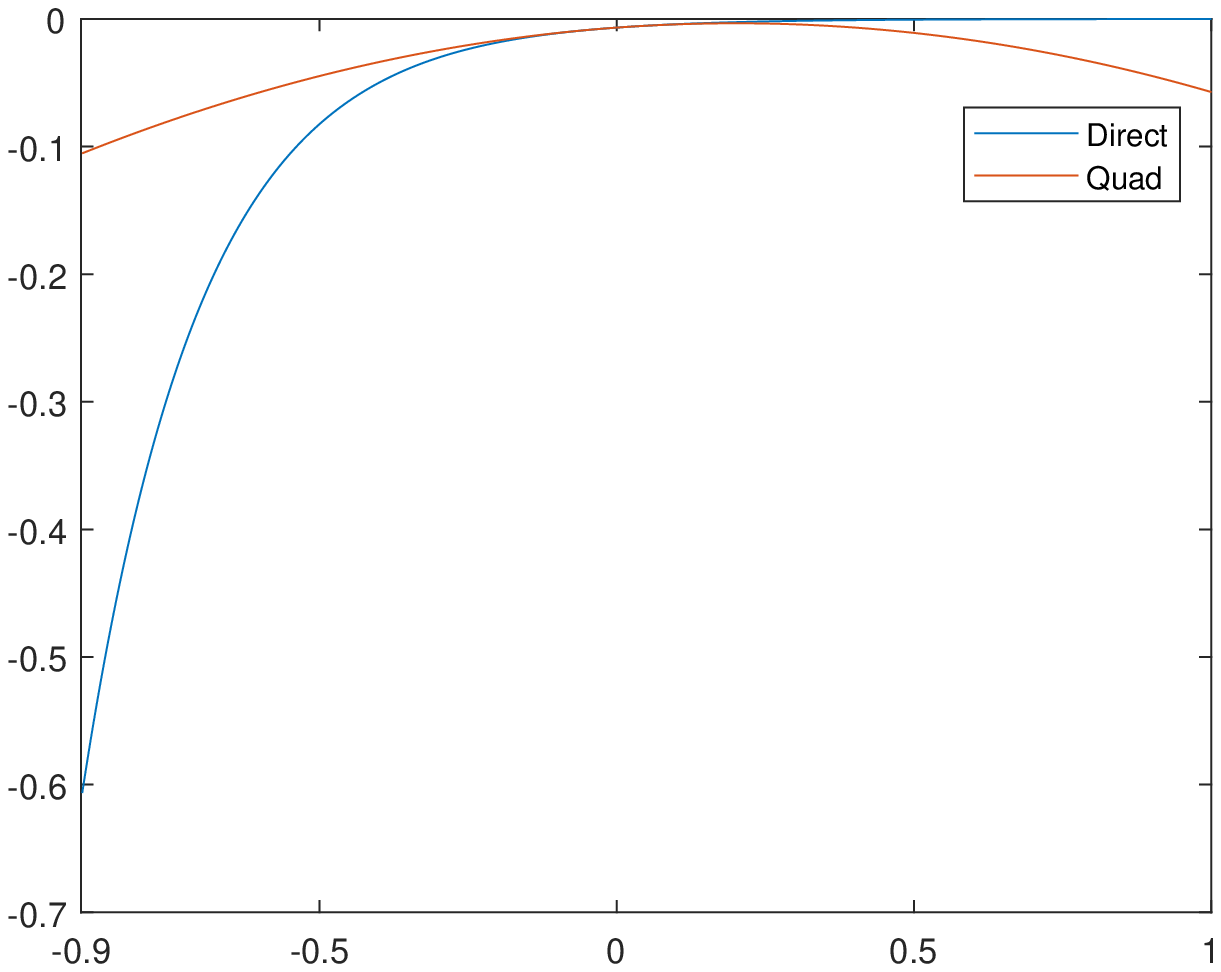}
			\caption{$a=5$}
		\end{subfigure}
	\end{figure}
	
	\newpage
	\begin{figure}[H]
		\ContinuedFloat
		\begin{subfigure}{0.45\textwidth}
			\includegraphics[width=\linewidth]{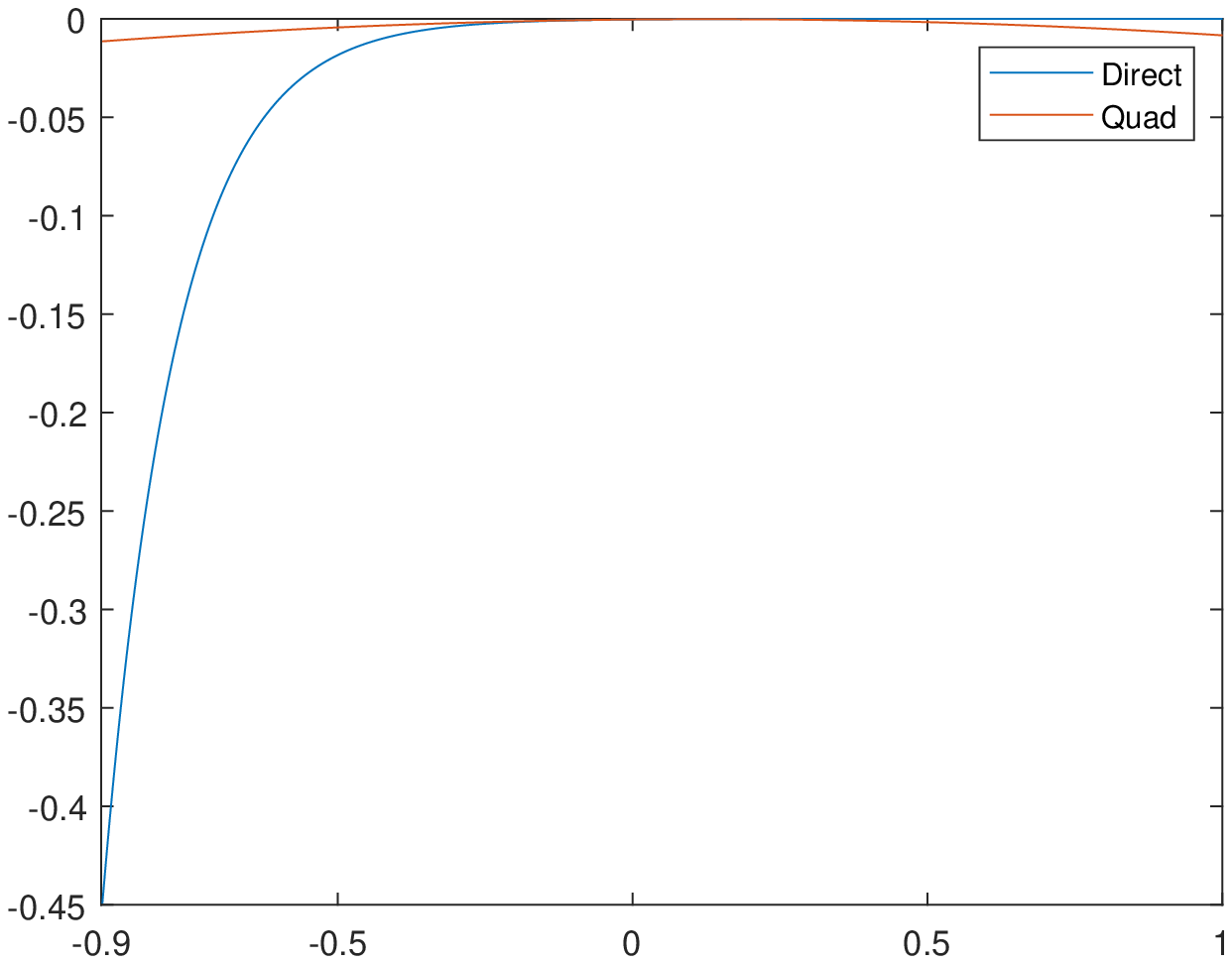}
			\caption{$a=8$}
		\end{subfigure}
		\hspace*{1in}
		\begin{subfigure}{0.45\textwidth}
			\includegraphics[width=\linewidth]{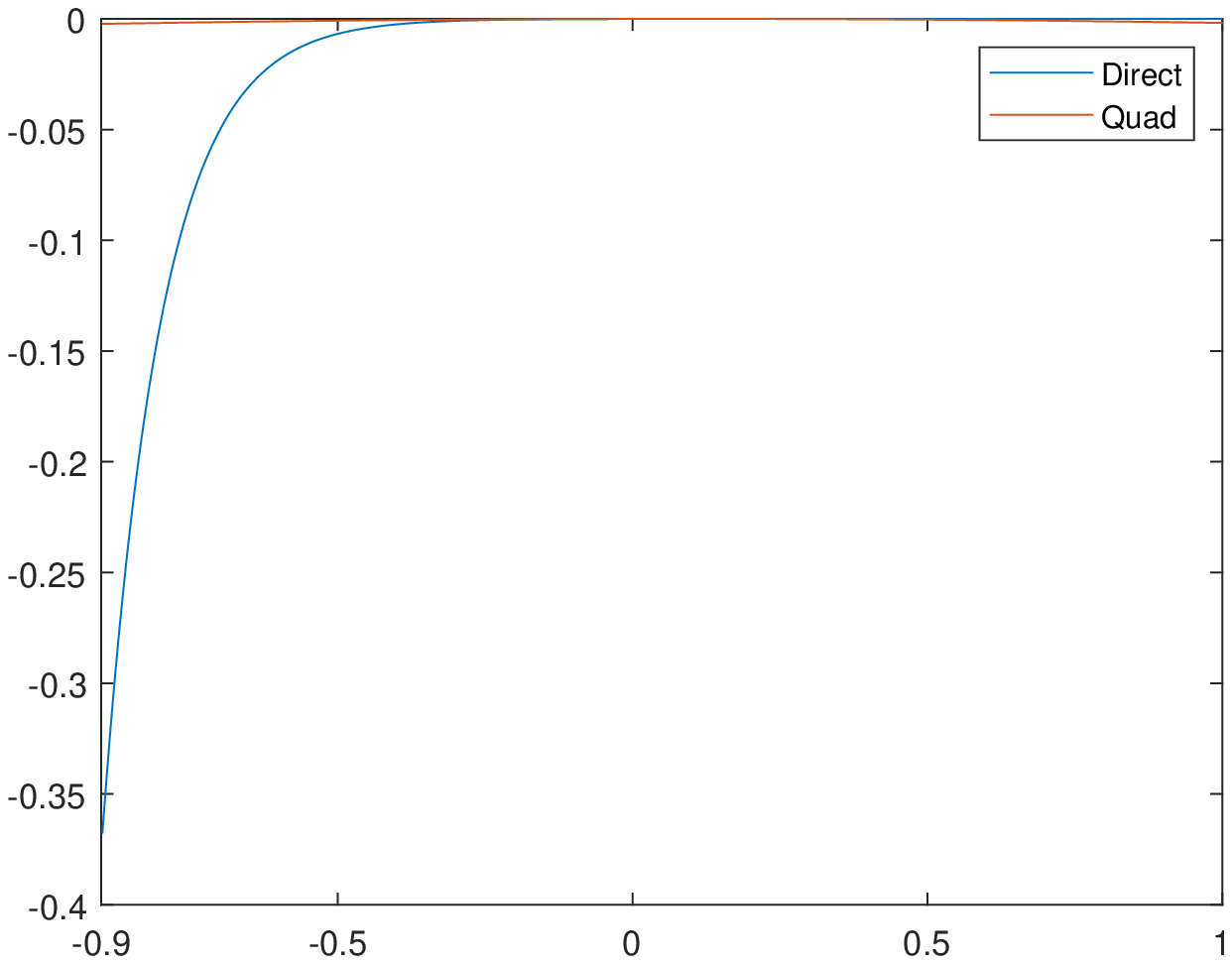}
			\caption{$a=10$}
		\end{subfigure}
		\begin{subfigure}{0.45\textwidth}
			\includegraphics[width=\linewidth]{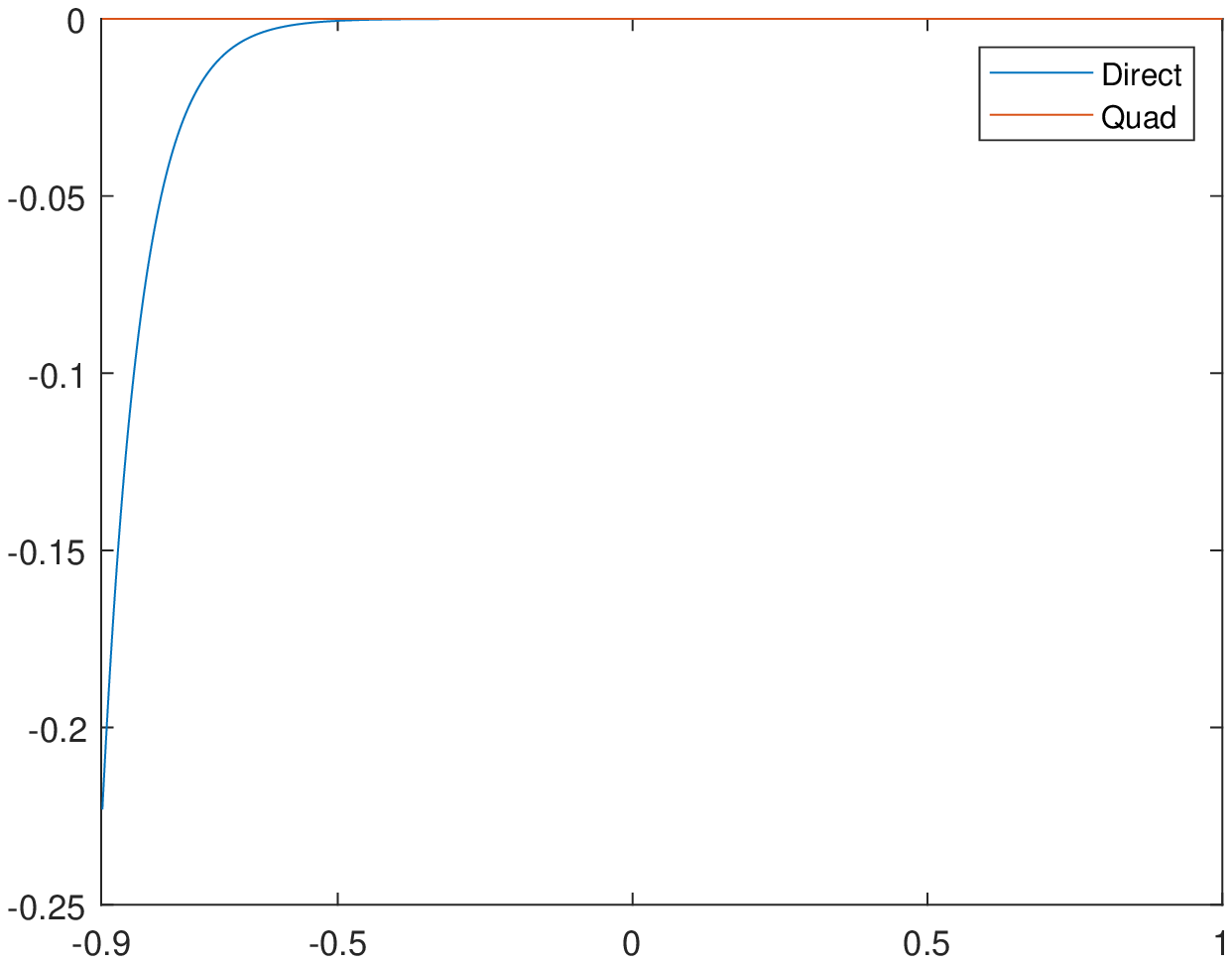}
			\caption{$a=15$}
		\end{subfigure}
		\hspace*{1in}
		\begin{subfigure}{0.45\textwidth}
			\includegraphics[width=\linewidth]{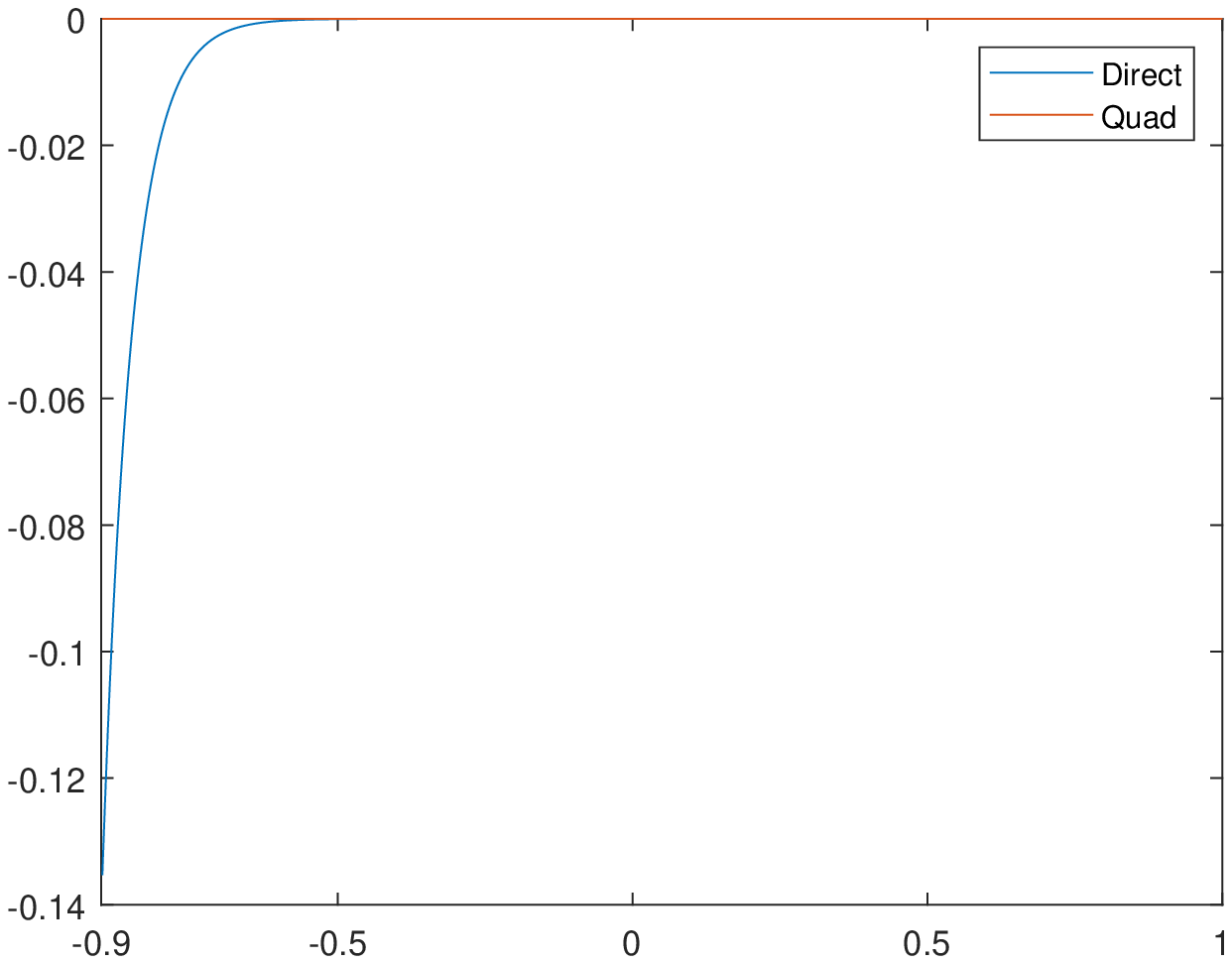}
			\caption{$a=20$}
		\end{subfigure}
	\end{figure}
	
	\newpage
	\begin{figure}[H]
		\caption{$-(1+Z)^{-a}$ vs its Quadratic approximation around $0$, for $Z\in[-0.9,1]$}
		\begin{subfigure}{0.45\textwidth}	
			\includegraphics[width=\linewidth]{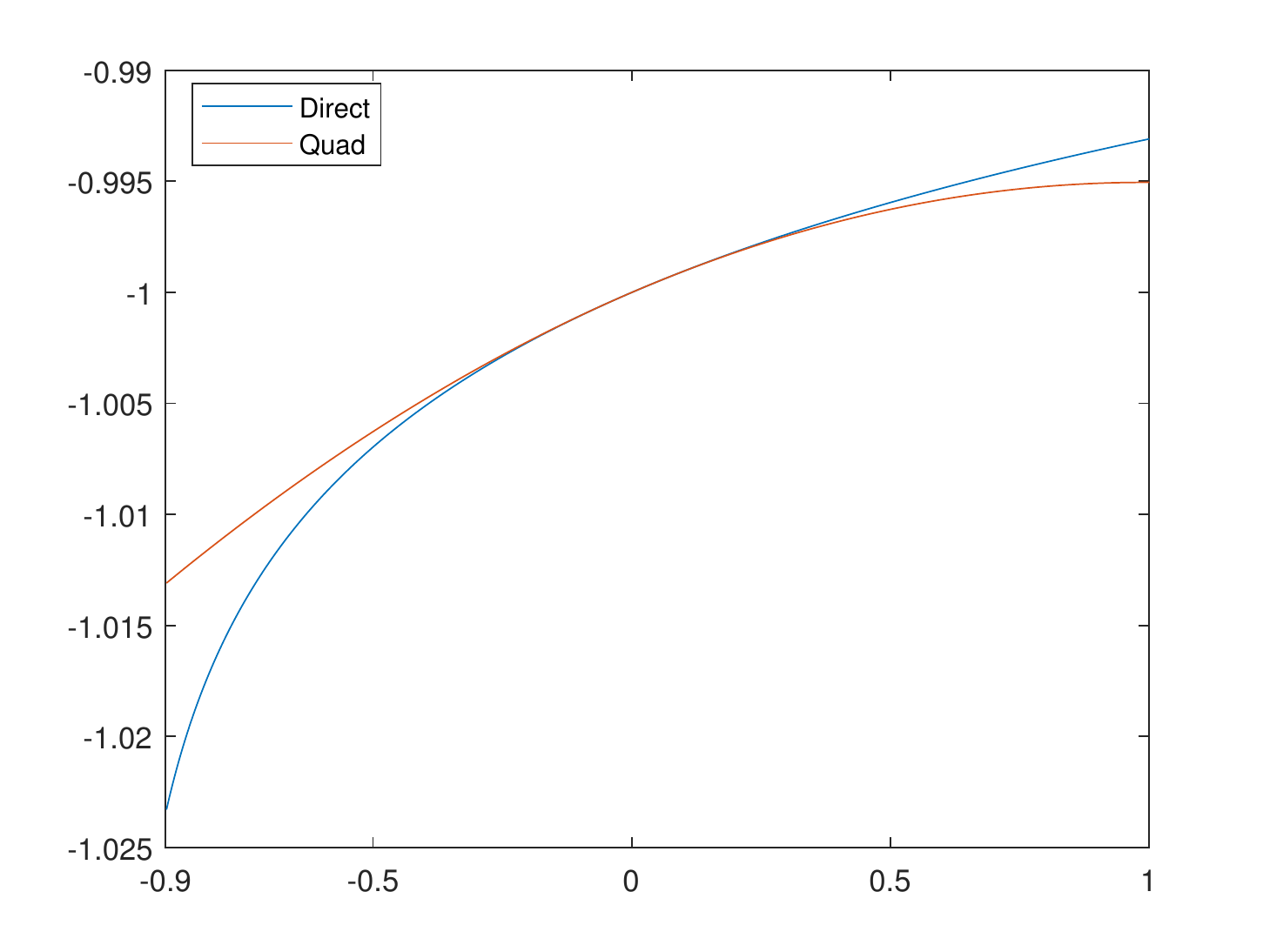}
			\caption{$a=0.01$}
		\end{subfigure}
		\hspace*{1in}
		\begin{subfigure}{0.45\textwidth}
			\includegraphics[width=\linewidth]{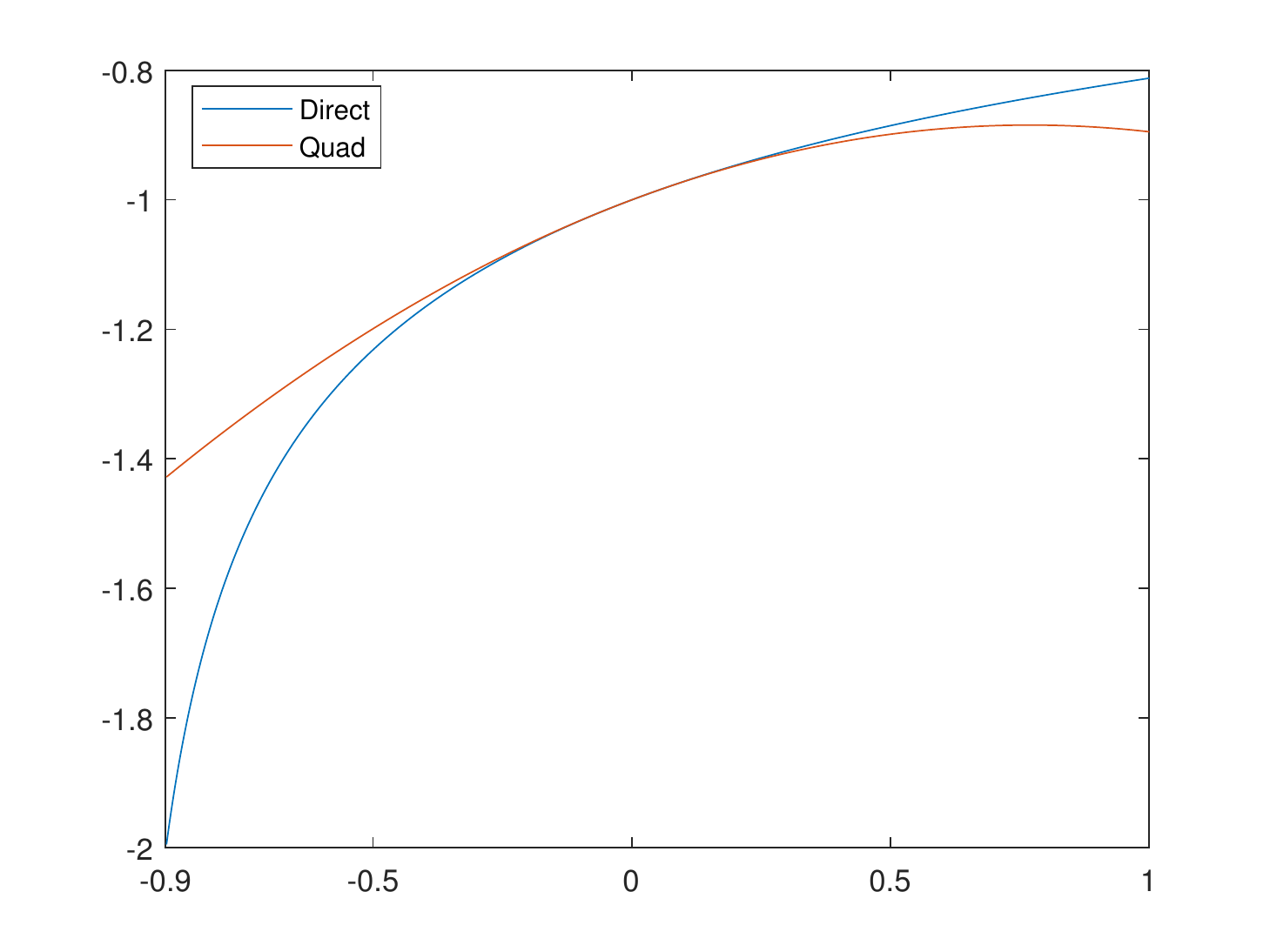}
			\caption{$a=0.30$}
		\end{subfigure}
		\begin{subfigure}{0.45\textwidth}
			\includegraphics[width=\linewidth]{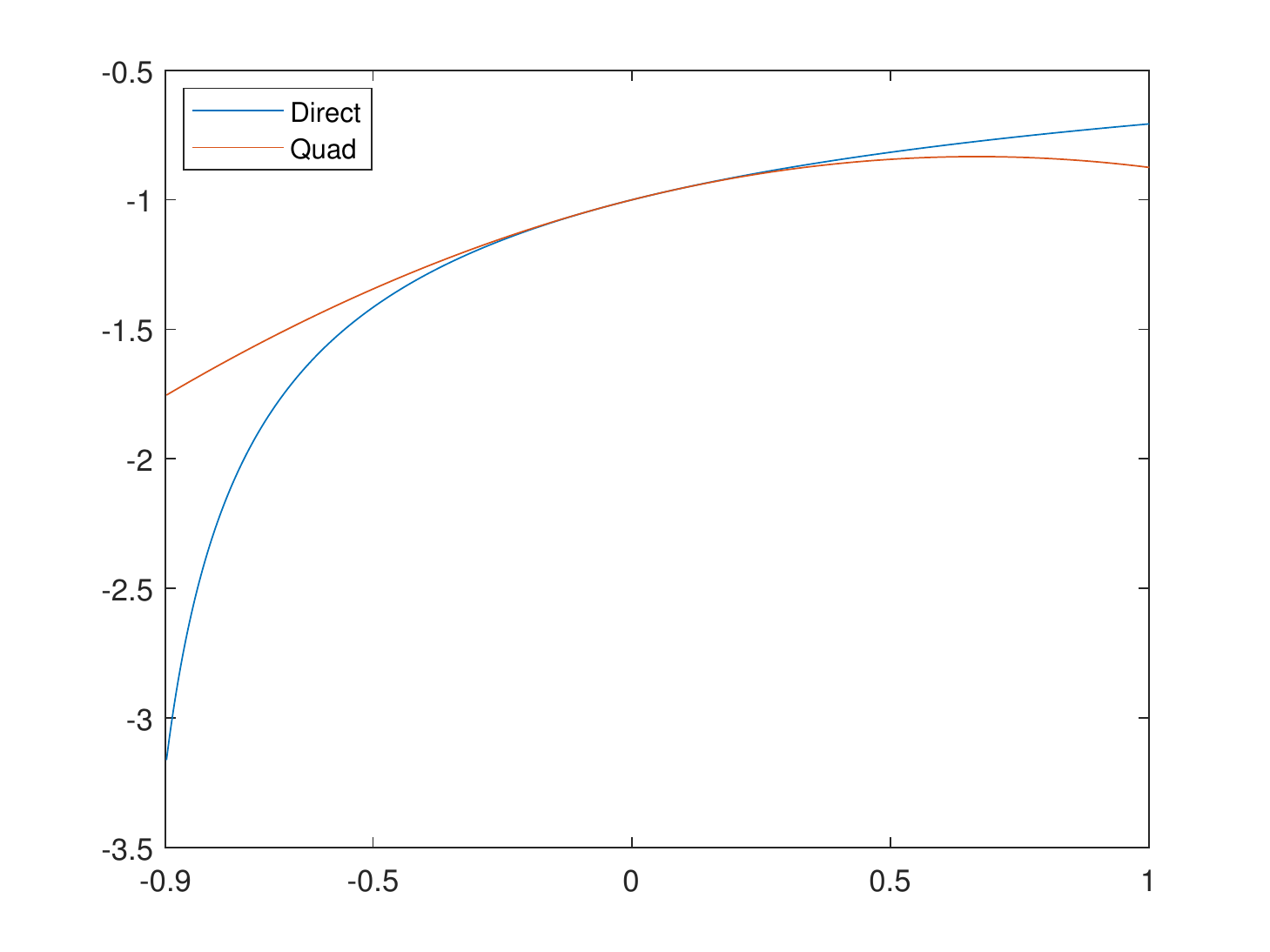}
			\caption{$a=0.50$}
		\end{subfigure}
		\hspace*{1in}
		\begin{subfigure}{0.45\textwidth}
			\includegraphics[width=\linewidth]{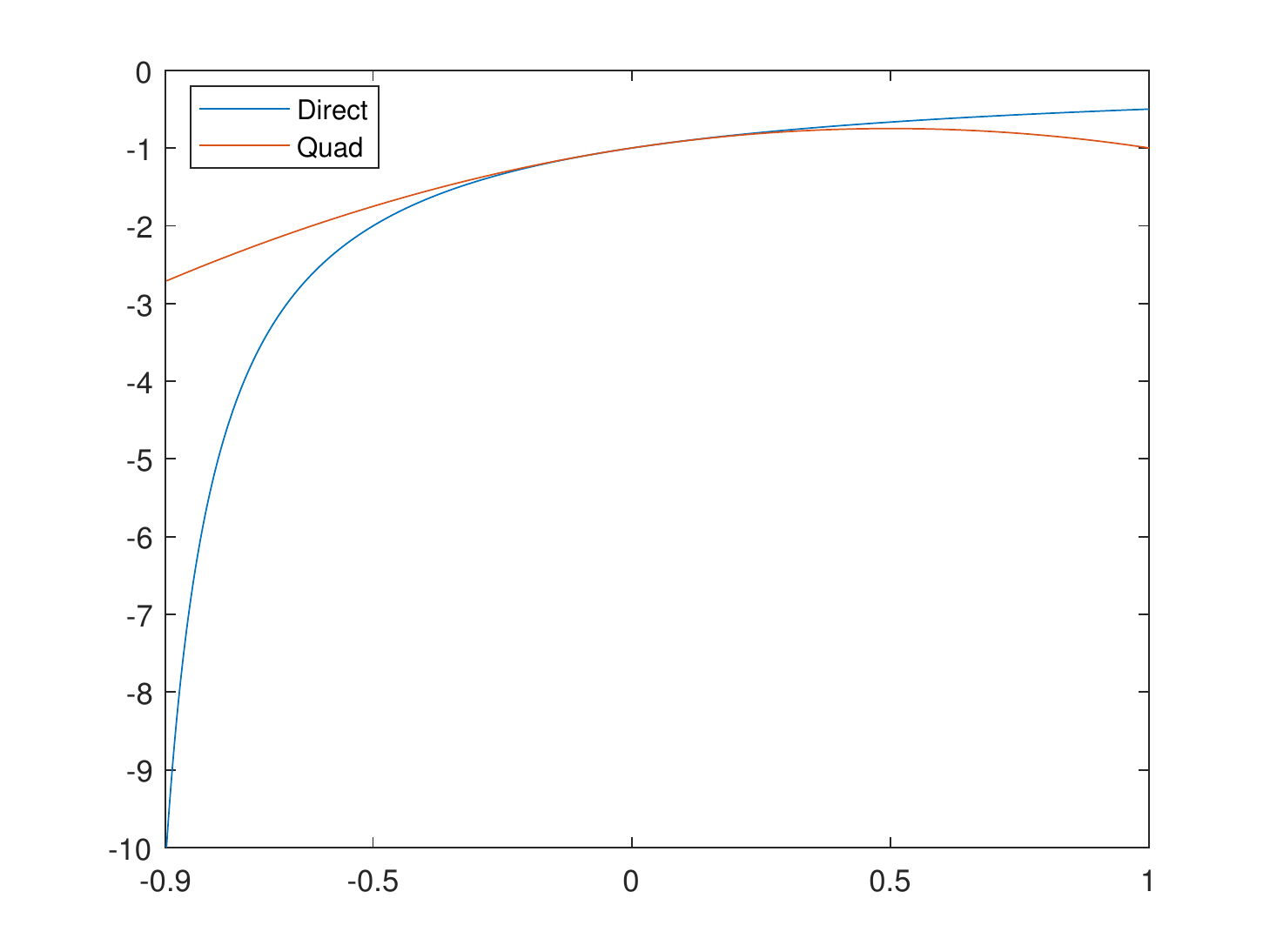}
			\caption{$a=1$}
		\end{subfigure}
	\end{figure}
	
	\newpage
	\begin{figure}[H]
		\ContinuedFloat
		\begin{subfigure}{0.45\textwidth}	
			\includegraphics[width=\linewidth]{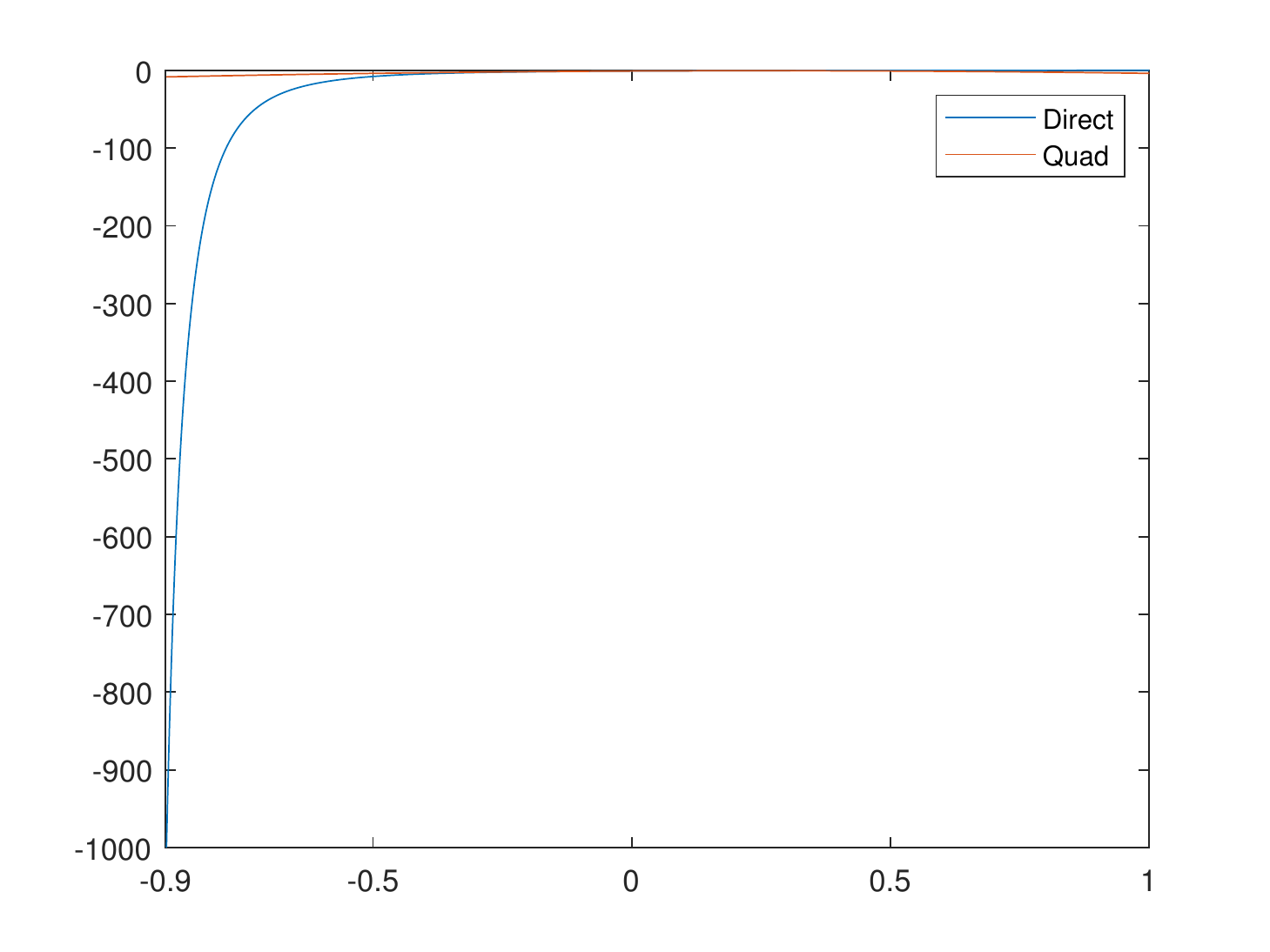}
			\caption{$a=3$}
		\end{subfigure}
		\hspace*{1in}
		\begin{subfigure}{0.45\textwidth}
			\includegraphics[width=\linewidth]{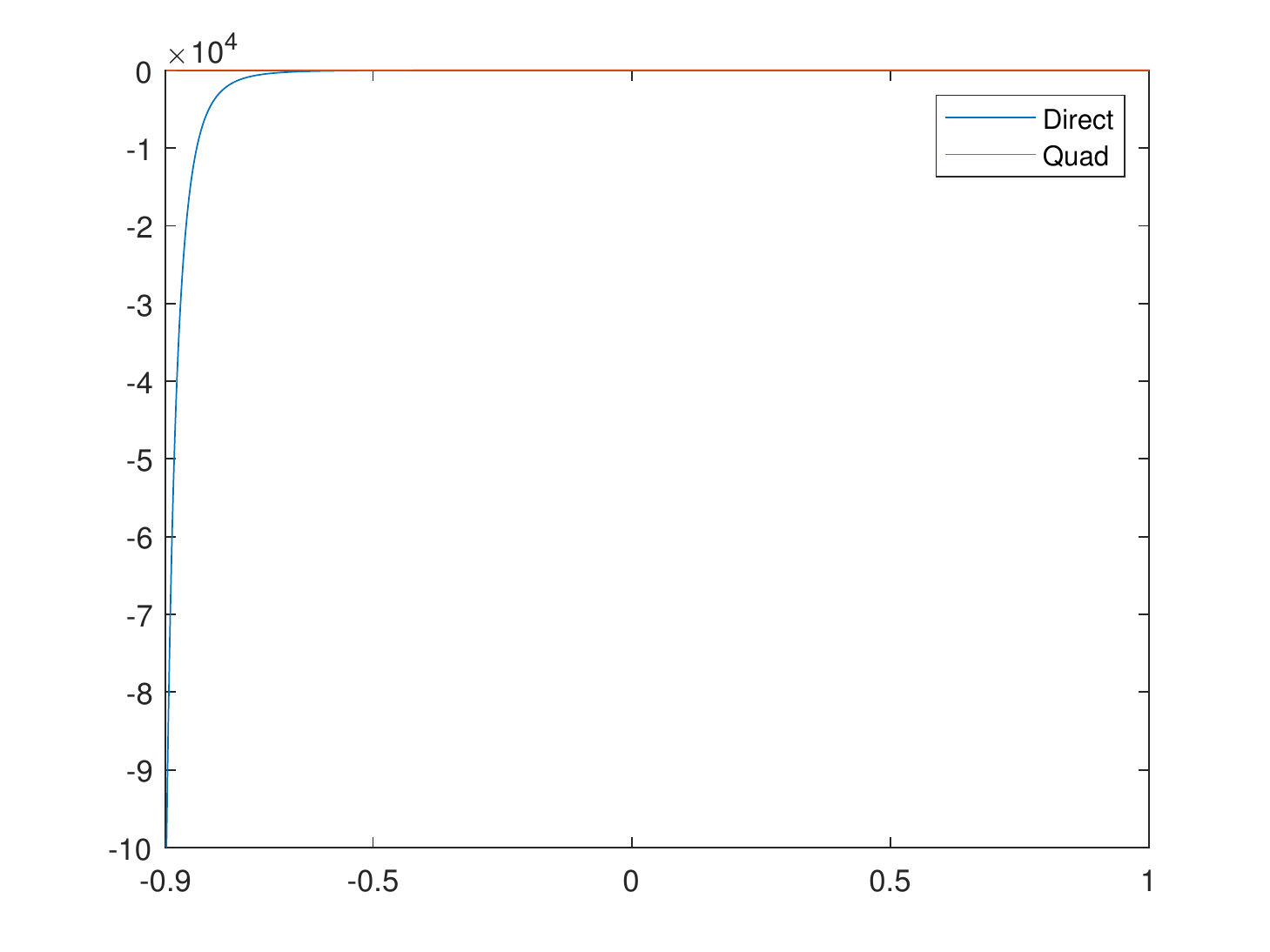}
			\caption{$a=5$}
		\end{subfigure}
		\begin{subfigure}{0.45\textwidth}
			\includegraphics[width=\linewidth]{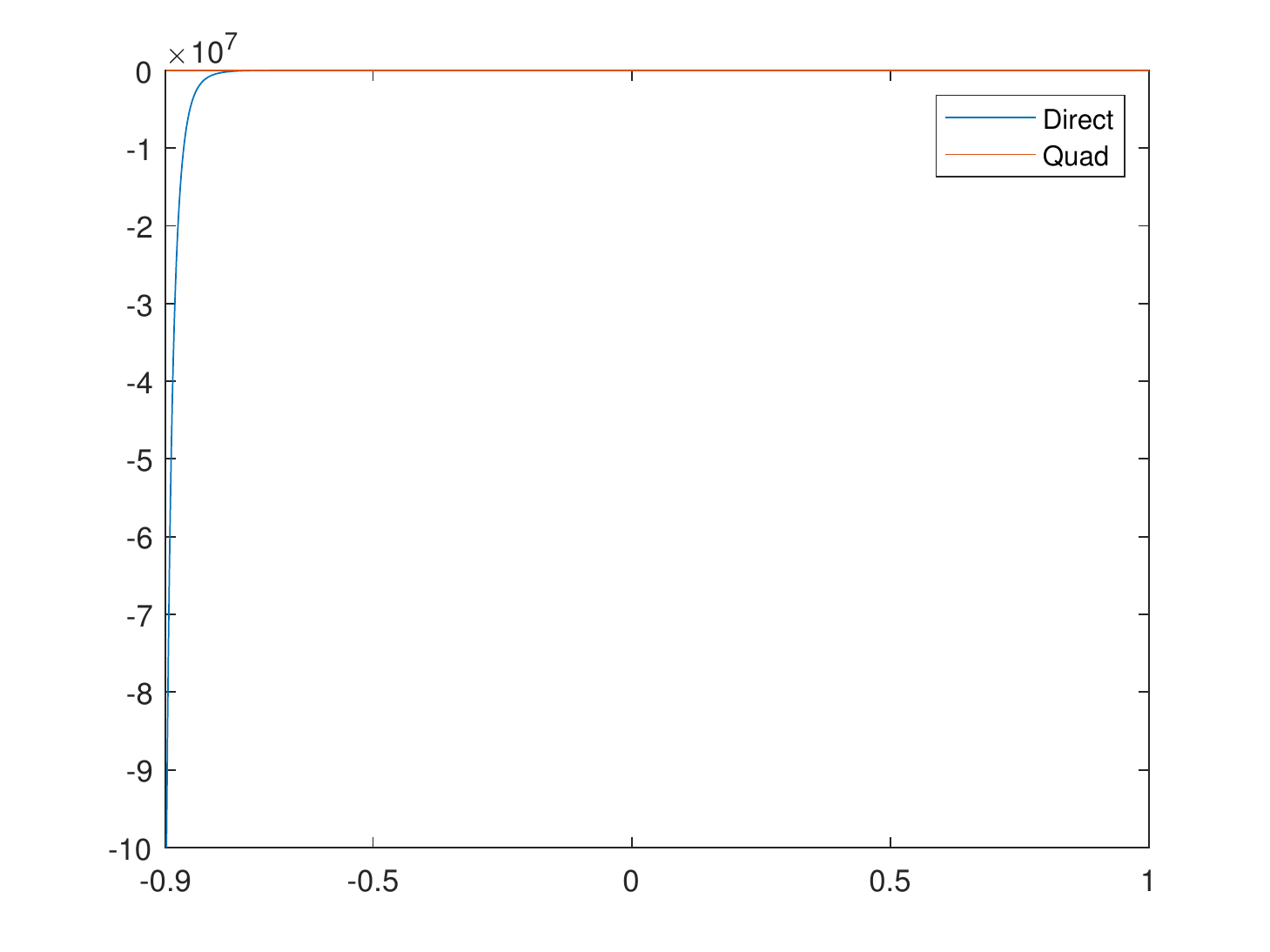}
			\caption{$a=8$}
		\end{subfigure}
		\hspace*{1in}
		\begin{subfigure}{0.45\textwidth}
			\includegraphics[width=\linewidth]{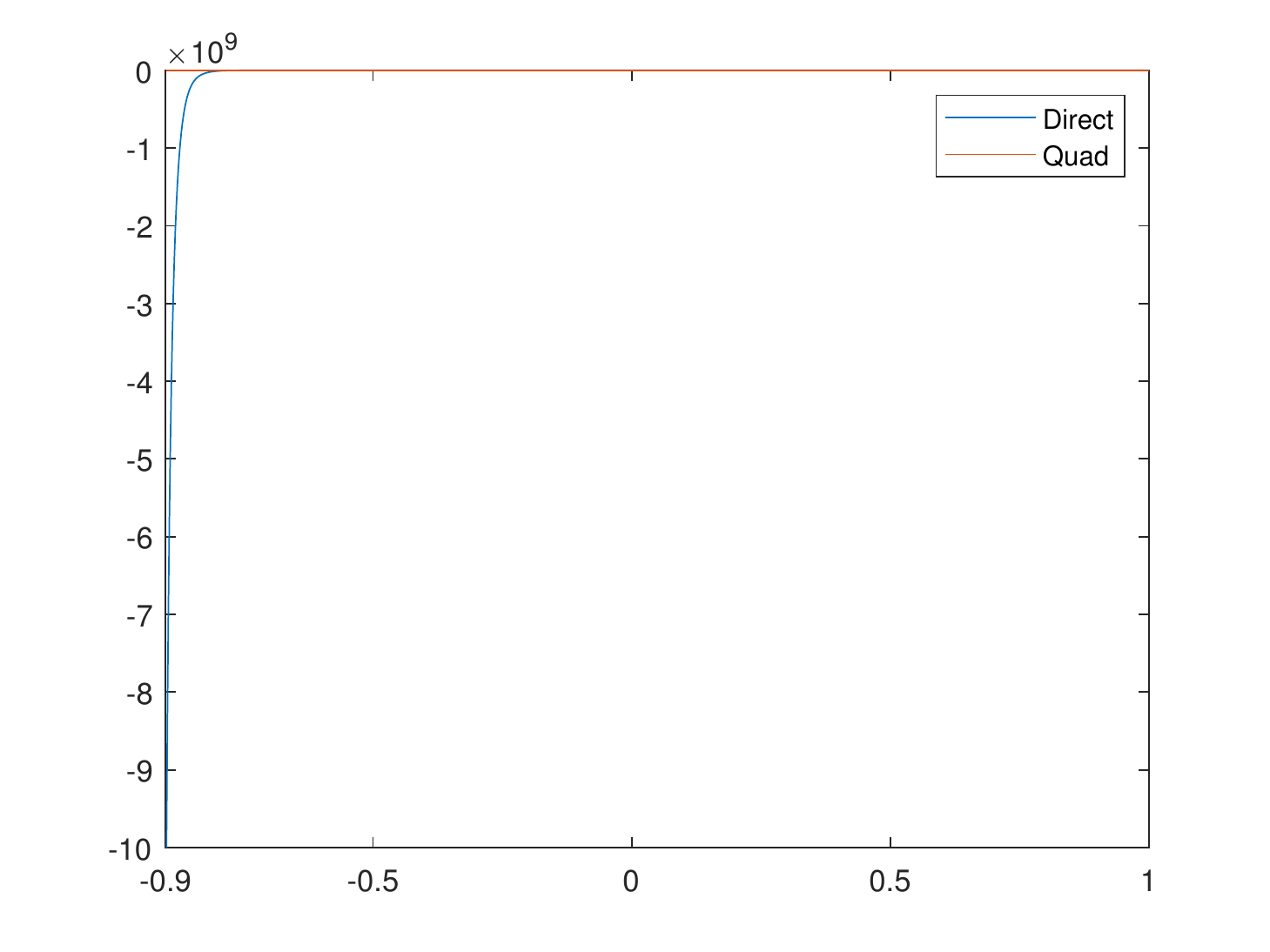}
			\caption{$a=10$}
		\end{subfigure}
		\begin{subfigure}{0.45\textwidth}	
			\includegraphics[width=\linewidth]{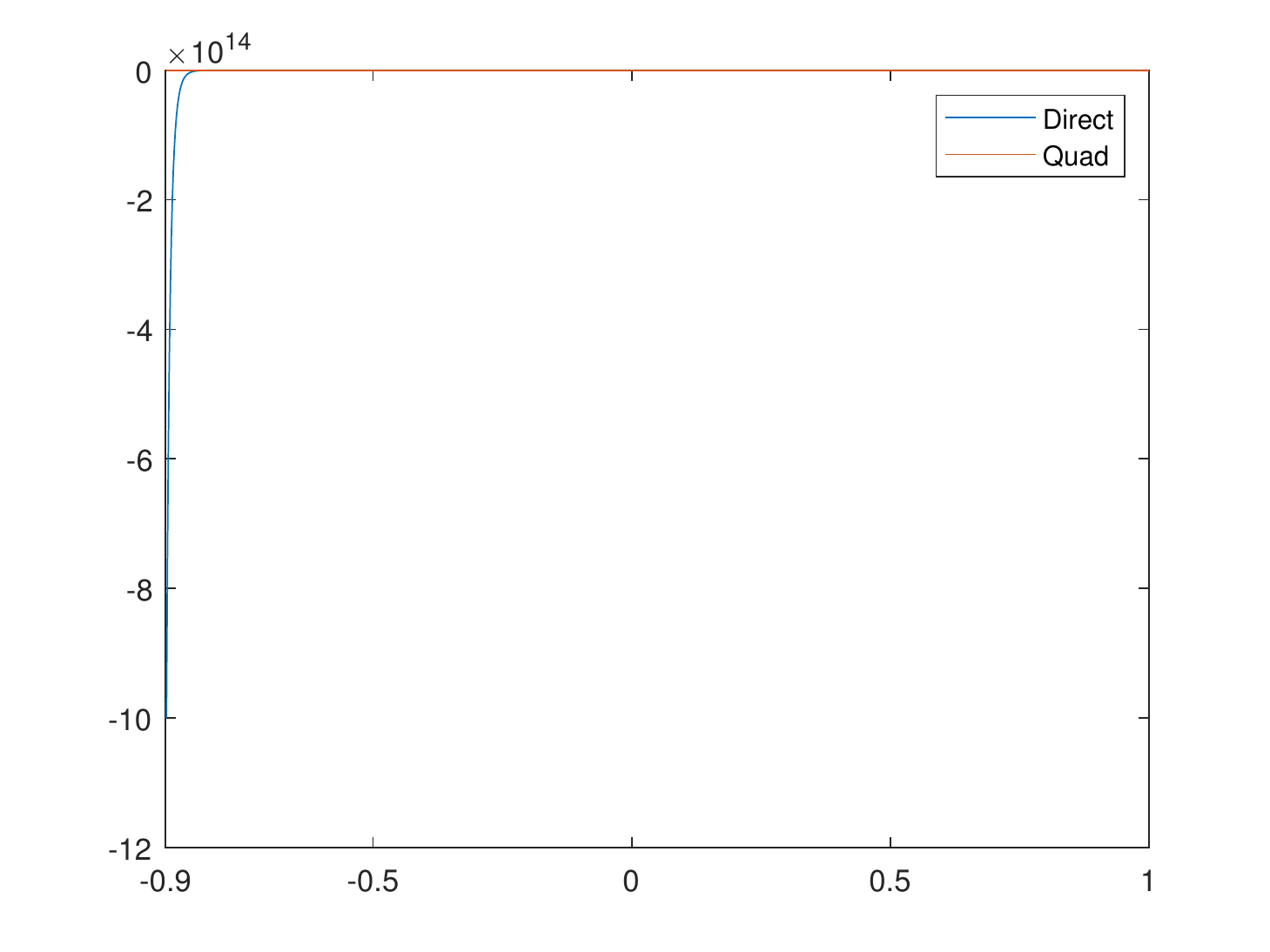}
			\caption{$a=15$}
		\end{subfigure}
		\hspace*{1in}
		\begin{subfigure}{0.45\textwidth}
			\includegraphics[width=\linewidth]{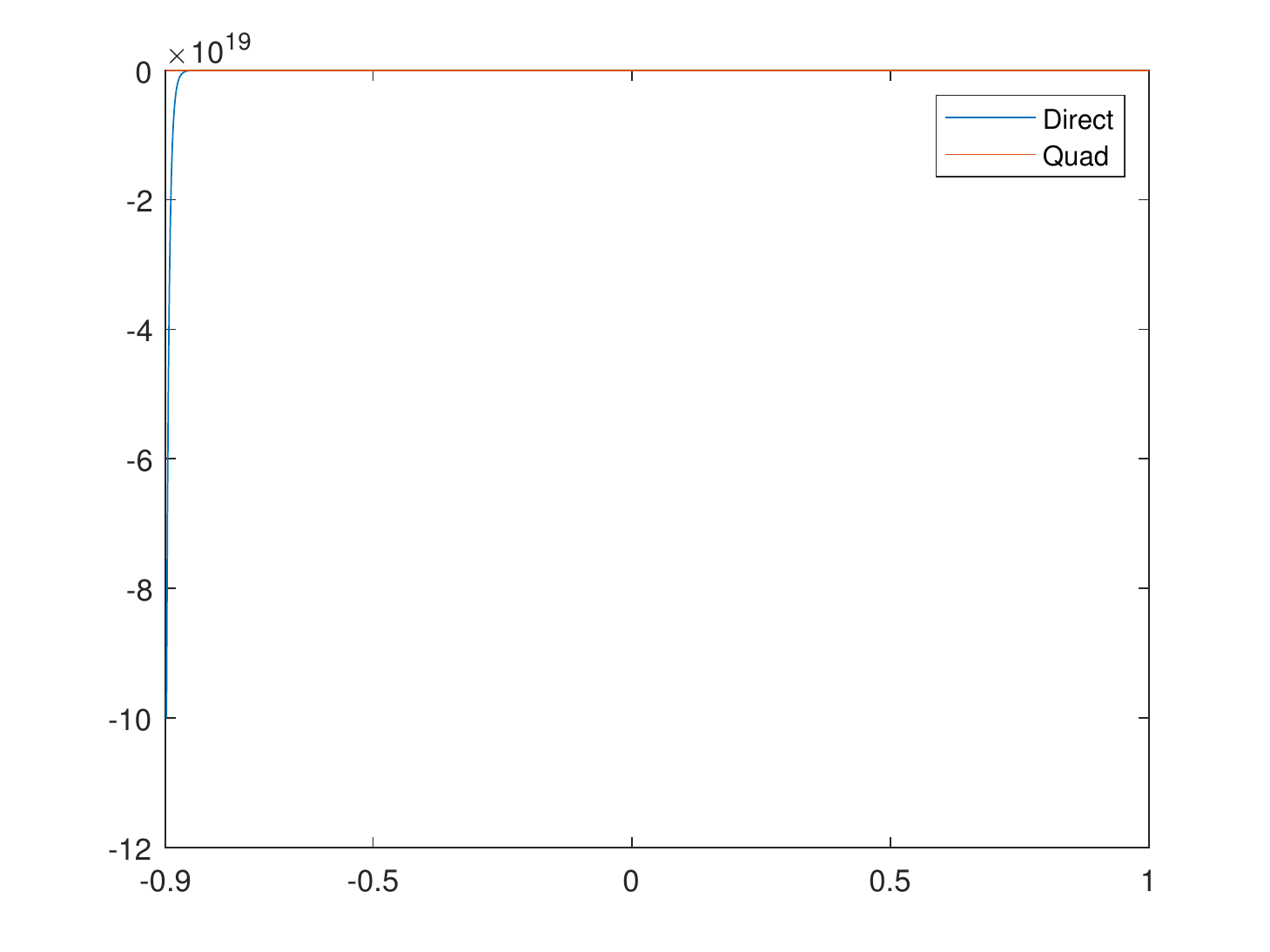}
			\caption{$a=20$}
		\end{subfigure}
	\end{figure}
	
	\newpage
	\printbibliography
	
\end{document}